\documentclass[aps,pra,
               reprint,
               a4paper,
               amsmath,amssymb,superscriptaddress,showpacs]{revtex4-1}

\usepackage[utf8]{inputenc}
\usepackage[english]{babel}
\usepackage{bm}
\usepackage{color}
\usepackage{multirow}
\usepackage{mathtools}
\usepackage{longtable}

\newcommand{\diag}{\mathop{\mathrm{diag}}\nolimits}
\newcommand{\vn}{\bm{\nabla}}
\newcommand{\ep}{\varepsilon_{0}}
\newcommand{\ee}{\varepsilon}
\newcommand{\pt}{\partial_{t}}

\newcommand{\vk}{\varkappa}

\begin{document}

\title{Optical chirality in gyrotropic media: symmetry approach}

\author{Igor \surname{Proskurin}}
\email{iprosk@ouj.ac.jp}
\affiliation{Faculty of Science, Hiroshima University, 
             Higashi-Hiroshima, Hiroshima 739-8526, Japan}
\affiliation{Institute of Natural Sciences, Ural Federal University,
             Ekaterinburg 620002, Russia}
             
\author{Alexander S. \surname{Ovchinnikov}}
\affiliation{Institute of Natural Sciences, Ural Federal University,
             Ekaterinburg 620002, Russia}
\affiliation{Institute for Metal Physics, RAS, 620137, 
	         Ekaterinburg, Russia}
             
\author{Pavel \surname{Nosov}}
\affiliation{Institute of Natural Sciences, Ural Federal University,
             Ekaterinburg 620002, Russia}
             
\author{Jun-ichiro \surname{Kishine}}
\affiliation{Division of Natural and Environmental Sciences, 
             The Open University of Japan,
             Chiba 261-8586, Japan}
             
\begin{abstract}
We discuss optical chirality in different types of gyrotropic media. Our analysis is based on the formalism of nongeometric symmetries of Maxwell's equations in vacuum generalized to material media with given constituent relations. This approach enables us to derive directly conservation laws related to the nongeometric symmetries. For isotropic chiral media, we demonstrate that likewise free electromagnetic field, both duality and helicity generators belong to the basis set of nongeometric symmetries that guarantees the conservation of optical chirality. In gyrotropic crystals, which exhibit natural optical activity, the situation is quite different from the case of isotropic media. For light propagating along certain crystallographic direction, there arise two distinct cases, i.~e., (1) the duality is broken but the helicity is preserved, or (2) only the duality symmetry survives. We show that the existence of one of these symmetries (duality or helicity) is enough to define optical chirality. In addition, we present examples of low-symmetry media, where optical chirality can not be defined.
\end{abstract}
             
\maketitle

\section{Introduction}
\label{sec:intro}
The notion of chirality, the term originally coined by Lord Kelvin for
any object which cannot be superimposed onto its mirror image
\cite{Kelvin1904}, is perhaps one of the most fundamental concepts in
nature ranging from particle physics to biology \cite{Dreiling2014}.  
After the discovery of a new conservation law for Maxwell's equations in vacuum by Lipkin \cite{Lipkin1964}, it
was realized that not only material objects but also fields can be characterized by certain chirality \cite{Morgan1964, Kibble1964}.  In particular,  we can construct the conserving pseudoscalar for free electromagnetic field
\begin{equation}
  \label{eq:1}
  C_{\chi} = \frac{1}{2}\int d^{3} r  \left ( 
    \varepsilon_{0} \bm{E} \cdot \bm{\nabla} \times \bm{E} + \mu_{0}^{-1}\bm{B} \cdot \bm{\nabla} \times \bm{B} \right ),
\end{equation}
which is even under time-reversal ($T$) and odd under spatial inversion ($P$) transformations. These symmetry properties are consistent with definition of \emph{true chirality} proposed by Barron \cite{Barron1986, Barron2004}, who stressed that we should distinguish it from \emph{false chirality} with broken $T$-symmetry. In this respect, this quantity is eligible to be called optical chirality, which was originally coined \emph{zilch} by Lipkin \cite{Lipkin1964}.

Later it was realized that besides classical conservation laws arising from the invariance under the space-time Poincar\'{e} group, free electromagnetic field is invariant under the eight-dimensional Lie algebra of nongeometric symmetry transformations.  This algebra results in an infinite number of integro-differential conservation laws, which include zilch as a particular case \cite{Fushchich1987}.
Together with the approach based on the nongeometric symmetries \cite{Fushchich1987}, these conservation laws were studied by means of the Lie theory and the Noether theorem \cite{Krivskii1989a, Krivskii1989b, Ibragimov2008,Cameron2012, Bliokh2013, Cameron2014}, by using the analogy between Maxwell's and Dirac equations \cite{Barnett2014}, and through the gauge symmetry \cite{Philbin2013}.

For free electromagnetic field, it was demonstrated that the existence of the duality symmetry for Maxwell's equations, i.e. the linear transformation that mixes electric and magnetic fields, engenders automatically the preservation of optical helicity,  i.e. the projection of total angular momentum on the direction of linear momentum \cite{Calkin1965, Zwanziger1968, Drummond1999, Drummond2006}.

The renewed interest to optical chirality has been stimulated through interdisciplinary studies in molecules and metamaterials \cite{Tang2010, Hendry2010, Tang2011, Hendry2012, Kamenetskii2013, Tomita2014}.  Notable progress in the field of optical angular momentum is also relevant to this directions \cite{Yao2011, Bliokh2015}. The relationship between optical chirality, helicity, and spin angular momentum of light was discussed in a number of papers \cite{Afanasiev2007, Bliokh2011, Coles2012, Cameron2012a, Barnett2012, Bliokh2014,Bliokh2014a,Bliokh2015a,Bliokh2016}.  A recent review on the general space-time symmetries of the Maxwell's equations can be found in \cite{Dressel2015}.

For electromagnetic field in media, the problem of optical chirality  has been considered in Refs.~\cite{Philbin2013, Ragusa1994, Ragusa1996}.  In dispersive media, the generalization of Lipkin's zilch  was established by Philbin \cite{Philbin2013} by means of the Noether's theorem applied to the specific gauge transformation of the magnetic vector potential. In isotropic chiral media, first order electromagnetic conservation laws were heuristically constructed by Ragusa in the relativistically noncovariant \cite{Ragusa1994} and covariant forms \cite{Ragusa1996}.

The purpose of this paper is to develop a systematic approach to the optical chirality in 
gyrotropic media.  To provide a theoretical basis for our treatment,  we invoke the formalism of nongeometric symmetries in vacuum \cite{Fushchich1987}, and generalize it to media taking heed of the corresponding constituent equations.  The key idea is to find the invariance algebra of nongeometric symmetries in media and to investigate whether the basis set of this algebra includes the transformations of duality and helicity.

At first, we analyze nongeometric symmetries in isotropic chiral media.  Likewise electromagnetic field in vacuum \cite{Calkin1965, Zwanziger1968}, isotropic chiral medium is self-dual, which automatically means that helicty is preserved \cite{Fernandez2013}. However, as we explicitly show, in chiral media, the original eight-dimensional invariance algebra of free electromagnetic field is broken down to  its four-dimensional subalgebra due to lack of the inversion symmetry. Duality and helicity are two essential generators of this subalgebra. For spatially nonuniform chiral medium, we find that the actual expression for optical chirality depends on the choice of constituent relations, in order to guarantee continuity of the chirality flow.

Next, we consider optical chirality in gyrotropic crystals, where surrounding symmetry is more restrictive, and, globally, neither duality nor helicity symmetry transformations are allowed. Although these symmetries can survive along the principal crystalline axes, the equivalence between duality and helicity, that holds in isotropic media,  is lost. We find that in crystals with gyrotropic birefrigence both duality and helicity operations are allowed along the principal axis. Crystals possessing natural optical activity and belonging to the point groups $C_n$ or $C_{nv}$ ($n\geq 3$) provide an example, where helicity is preserved along the principal axis \cite{Fernandez2013a}, while the duality symmetry is broken. The opposite situation is realized in achiral materials with natural optical activity, where the duality symmetry along the principal axis is preserved without helicity. The existence of either duality or helicity transformations along a certain direction leads to the conservations law for optical chirality, which is consistent with underlying symmetries.

In concluding remarks, we consider low-dimensional crystals, where we encounter the invariance algebra with neither  preserved duality nor preserved helicity, which makes optical chirality ill defined.


\section{Free electromagnetic field}
\label{sec:vac}
We begin with a brief review of the latest developments in symmetry analysis of the electromagnetic field in vacuum. We give a pedagogical introduction in the method of nongeometrical symmetries, which is generalized in the subsequent sections to electromagnetic field in media.

The discovery of a conservation law in Eq.~(\ref{eq:1}) stimulated the discussion of the related `hidden' symmetries for the electromagnetic field. Historically, it had been established shortly after the formulation of electrodynamics that the Maxwell's equations in free space
\begin{align}
\label{eq:Max3}
\vn \times \bm E = -\pt \bm B,   \qquad &
\vn \times \bm B =  \pt \bm E,   \\
\label{eq:Max4}
\vn \cdot \bm E = 0,             \qquad & 
\vn \cdot \bm B = 0,          
\end{align}
(in this section we use $c = 1$ and $\hbar = 1$) remain invariant under the duality transformation 
\begin{align} \label{d1}
\bm{E} &\to  \bm{E} \cos\theta + \bm{B}\sin\theta,\\ \label{d2}
\bm{B} &\to -\bm{E} \sin\theta + \bm{B}\cos\theta,
\end{align}
which can be viewed as `rotation' in the pseudo-space of  $ \bm{E} $ and $ \bm{B} $ vectors (for a review and historical background, see Refs.~\cite{Cameron2012,Bliokh2013}).

In contrast to the Maxwell's equations, standard Lagrangian formulation of electrodynamics is not symmetric under the duality transformation, and, to mitigate this obstacle, a duality symmetric form of the Lagrangian density was proposed \cite{Cameron2012,Bliokh2013}. The duality symmetric form, on the basis of the Noether theorem, ties up the transformation in Eqs.~(\ref{d1}, \ref{d2}) with conservation of optical helicity 
\begin{equation}\label{hel}
C_{\mathrm{hel}} = \frac{1}{2} \int d^3 r \left( \bm{A} \cdot \bm \nabla \times  \bm{A} + \bm{C} \cdot \bm \nabla \times \bm{C} \right),
\end{equation}
written in terms of magnetic ($ \bm{A} $) and electric ($ \bm{C} $) vector potentials, which are determined by $ \bm{E} = -\bm{\nabla} \times \bm{C} = -\pt \bm{A} $ and $ \bm{B} = \bm{\nabla} \times \bm{A} $, and satisfy the symmetry transformation in Eqs.~(\ref{d1}, \ref{d2}). 

With respect to the optical helicity, Lipkin's zilch in Eq.~(\ref{eq:1}) can be regarded as the next order term in the infinite hierarchy of higher order conserving zilches \cite{Kibble1964,Cameron2012a}. Although being different form optical helicity in general, for monochromatic fields, Lipkin's zilch becomes proportional to the helicity. Both quantities are determined by the difference between left and right polarized photon numbers \cite{Bliokh2011,Coles2012}. Tang and Cohen proposed to use Lipkin's $00$-zilch as a measure of \emph{optical chirality} in light-matter interactions \cite{Tang2010,Tang2011}.

Different approach to zilch conservation laws has been developed in Refs.~\cite{Krivskii1989a,Krivskii1989b,Ibragimov2008} on the basis of the Lie-Noether analysis. Recently, Philbin explicitly demonstrated that zilch conservation can be obtained from the standard electromagnetic Lagrangian $ L = (E^2 - B^2)/2 $ by applying a specific `hidden' gauge symmetry transformation of the magnetic vector potential, $ \bm{A} \to \bm{A} + \eta \bm{\nabla} \times \pt \bm{A} $, with infinitesimal parameter $\eta$ \cite{Philbin2013}, similar to Calkin's original arguments \cite{Calkin1965}. 

Another powerful tool, different form the Lagrangian-Noether approach, that we used throughout this paper, is a method of the nongeometric symmetries developed by Fushchich and Nikitin \cite{Fushchich1987}. The advantage of this method is that it is solely based on the analysis of the equations of motion and, therefore, does not rely on any ambiguity in specific gauge choice or Lagrangian representation. This fact makes it possible to generalize this approach to Maxwell's equations in media with given constituent relations.

For the symmetry analysis, it is essential to find convenient representation of the Maxwell's equations.
We use Silberstein-Bateman form, which is convenient to work in the momentum space. The transformation to the momentum space is reached by
\begin{align}
\bm{E}(t,\bm{r}) &= \frac{1}{(2\pi)^{3/2}} \int d^{3} p \,
e^{i \bm{p} \cdot \bm{r}}  \bm{E}(t,\bm{p}), \\
\bm{B}(t,\bm{r}) &= \frac{1}{(2\pi)^{3/2}} \int d^{3} p \,
e^{i \bm{p} \cdot \bm{r}}  \bm{B}(t,\bm{p}).
\end{align}

In the Silberstein-Bateman representation, the first pair of the Maxwell's equations in Eq.~(\ref{eq:Max3})
is expressed in terms of a Shroedinger-like equation for a six-component
vector column $\phi(t,\bm{p}) = (\bm{E}, \bm{B})^{T}$
\begin{equation}
  \label{eq:L1}
  i \frac{\partial \phi(t, \bm{p})}{\partial t} = \mathcal{H} \phi(t, \bm{p}),
\end{equation}
where $ \mathcal{H} $ is the Hermitian matrix given by
\begin{equation}
  \label{eq:Hamilt}
  \mathcal{H} = -\sigma_{2} \otimes (\hat{\bm{S}} \cdot \bm{p}) = 
  \begin{pmatrix}
  	0                              & i(\hat{\bm{S}} \cdot \bm{p}) \\
  	-i(\hat{\bm{S}} \cdot \bm{p}) & 0
  \end{pmatrix},
\end{equation}
which can be considered as an analogue of the quantum-mechanical Hamiltonian. Here, we introduced $3 \times 3$ spin matrices $\hat S_\alpha$ with matrix elements $(\hat{S}_{\alpha})_{\beta \gamma} = -i \epsilon_{\alpha \beta \gamma}$, where $\epsilon_{\alpha\beta\gamma}$ is the Levi-Civita
symbol, $\otimes$ means the Cartesian product, and $\sigma_{\mu}$ ($\mu =
1,2,3$) are $2\times2$ Pauli matrices.
In what follows, we hold the following notations. The
`hat' is used to distinguish $3 \times 3$ matrices. Calligraphic style is reserved for the $6 \times 6$ matrices. $\sigma_{0}$ and $\hat{I}$ denote two- and three-dimensional unit matrices,
and Greek indices run over the three dimensional space.

The second pair of Maxwell's equations (\ref{eq:Max4}) is equivalent to the
additional constrain imposed on $\phi(t,\bm{p})$ \cite{Fushchich1987}
\begin{equation}
  \label{eq:L3}
  \mathcal{L}\phi(t, \bm{p}) = 0, \qquad {\cal L} = p^2 - (\bm{\mathcal{S}} \cdot \bm{p})^{2},
\end{equation}
where $\bm{\mathcal{S}} \equiv \sigma_{0} \otimes \hat{\bm{S}}$, which accounts for the transversal character of the electromagnetic field. For real $\phi(t,\bm{p})$, one should also require $ \phi^{*}(t,\bm{p}) = \phi(t,-\bm{p}) $.

Now let us find all the transformations in the $\bm{p}$-space given by matrices $\mathcal{Q}_{A}(\bm{p})$ that transform a solution of the Maxwell's equations $ \phi(t,\bm{p}) $ into another solution $\phi'(t, \bm p) = \mathcal{Q}_{A}(\bm p)\phi(t, \bm p)$. Following Ref.~\cite{Fushchich1987}, we will call $\mathcal{Q}_{A}(\bm{p})$  \emph{nongeometric symmetry} transformations.  The total number of such $ \mathcal{Q}_{A}(\bm p) $ is given in the theorem, which claims that the Maxwell's equations in vacuum are invariant under the eight-dimensional Lie algebra			
  \begin{align}
    \label{eq:QA}
    \mathcal{Q}_{1} &=   \sigma_{3} \otimes (\hat{\bm S} \cdot \tilde{\bm p}) \hat{D}, \qquad &
    \mathcal{Q}_{2} &= i \sigma_{2} \otimes \hat{I},                               \\
    \mathcal{Q}_{3} &= - \sigma_{1} \otimes (\hat{\bm S} \cdot \tilde{\bm p}) \hat{D}, \qquad &
    \mathcal{Q}_{4} &= - \sigma_{1} \otimes \hat{D},                               \\
    \mathcal{Q}_{5} &=   \sigma_{0} \otimes (\hat{
    \bm S} \cdot \tilde{\bm p}),         \qquad &
    \mathcal{Q}_{6} &= - \sigma_{3} \otimes \hat{D},                               \\
    \mathcal{Q}_{7} &=   \sigma_{0} \otimes \hat{I},                               \qquad & 
    \mathcal{Q}_{8} &= i \sigma_{2} \otimes (\hat{\bm S} \cdot \tilde{\bm p}),
    \label{eq:Q8}
  \end{align}
  where $\tilde{\bm p} = \bm p/p$, $\hat{D} = \hat{D}_{0} + \hat{D}_{1}$
  with $(\hat{D}_{0})_{\alpha \beta} = (\delta -f)p_{\alpha}p_{\beta}/(p^2 \delta)$
  and 
  \begin{equation}
  \hat{D}_{1} = \frac{1}{\delta} \left(
  \begin{array}{ccc}
  f - 2 p_{2}^{2} p_{3}^{2}                                       & 
  p_{1} p_{2} p_{3}^{2}                                           & 
  p_{1} p_{3} p_{2}^{2}                                           \\
  p_{1} p_{2} p_{3}^{2}                                           &
  f - 2 p_{1}^{2} p_{3}^{2}                                       &
  p_{2} p_{3} p_{1}^{2}                                           \\
  p_{1} p_{3} p_{2}^{2}                                           &
  p_{2} p_{3} p_{1}^{2}                                           &
  f - 2 p_{1}^{2} p_{2}^{2}                                        
  \end{array}
  \right),
  \end{equation}
where $f=p_{1}^{2} p_{2}^{2} + p_{1}^{2} p_{3}^{2} + p_{2}^{2} p_{3}^{2}$ and
$\delta = [p_{1}^{4}(p_{2}^{2} - p_{3}^{2})^{2} 
+ p_{2}^{2}p_{3}^{2}(p_{1}^{2} - p_{2}^{2})(p_{1}^{2} - p_{3}^{2})]^{1/2}$.
The basis elements form the algebra, which is isomorphic to the Lie algebra of the group $U(2) \otimes
U(2)$ \cite{Fushchich1987}. 

All the basis elements in Eqs.~(\ref{eq:QA})--(\ref{eq:Q8}) commute with $ \mathcal{H} $ in Eq.~(\ref{eq:Hamilt}) and act as generators for the continuous symmetry transformations 
\begin{equation} \label{sym}
\phi(t, \bm{p}) \to \exp(\mathcal{Q}_{A} \theta_{A})\phi(t, \bm{p}),
\end{equation}  
where $ \theta_{A} $ are real parameters.

Analogy with quantum mechanics suggests that one can find conserving quantities related to the symmetry transformations $ \mathcal{Q}_{A} $, which can be conveniently formulated in terms of bilinear forms
\begin{equation}
  \label{eq:IOM}
  \left \langle \mathcal{Q}_{A} \right \rangle = 
   \frac{1}{2}\int d^{3} p \, \phi^{\dag}(t,\bm{p})\, \mathcal{M} 
  \mathcal{Q}_{A}\, \phi(t,\bm{p}), 
\end{equation} 
where $\mathcal{M}$ is any operator commutative with $\mathcal{H}$.

Some symmetry transformations in Eqs.~(\ref{eq:QA})--(\ref{eq:Q8}) have transparent
physical interpretations. For example, $ \mathcal{Q}_{7} $ is the identity transformation. The 
corresponding conserving quantity is the electromagnetic energy 
\begin{equation}
\left \langle \mathcal{Q}_{7} \right \rangle = 
\frac{1}{2}\int d^{3} p  \phi^{\dag}(t,\bm{p})
\mathcal{Q}_{7} \phi(t,\bm{p}) = \frac{1}{2}\int d^{3} p \left( E^2 + B^2\right).
\end{equation} 

Duality transformation in Eqs.~(\ref{d1}, \ref{d2}) also belongs to the class on nongeometric symmetries. This symmetry is generated by $ \mathcal{Q}_{2} $, in accordance with Eq.~(\ref{sym}).

Electromagnetic chirality can be expressed as a conserving quantity that corresponds to the operator 
$ p\mathcal{Q}_{5} $. Indeed, the expression 
\begin{equation}
  \label{eq:Zvac}
  C_\chi =  \frac{1}{2} \int d^{3} p \, \phi^{\dag}(t,\bm{p})\, 
  (\bm{\mathcal{S}} \cdot \bm{p}) \, \phi(t,\bm{p}),
\end{equation}
is transformed into Eq.~(\ref{eq:1}) in the real space. In what follows, we would refer $\mathcal{Q}_{5}$ as a \emph{helicity operator}, since electromagnetic helicity in Eq.~(\ref{hel}) can be also expressed in terms of a bilinear form containing this operator acting in the space $ (\bm{A},\bm{C}) $. 

By noting that $ p \mathcal{Q}_{5} \mathcal{Q}_{2} = -i \mathcal{H} =\pt $, we can find an alternative form of the optical chirality expressed via the duality operator
\begin{equation}
  \label{eq:Zvac1}
  C_\chi =  -\frac{i}{2}\int d^3 p \, \phi^\dag(t,\bm p) \,
  {\cal Q}_2 \, \partial_t \phi(t,\bm p).
\end{equation}
However, as we show in the next sections, the identity above between $ \mathcal{Q}_{5} $, $ \mathcal{Q}_{2} $, and $ \mathcal{H} $ does not necessary hold in crystals, where it is possible that either $ \mathcal{Q}_{2} $ or  $ \mathcal{Q}_{5} $ is allowed symmetry along certain crystallographic direction, but not both simultaneously.

Using the ambiguity in the choice of $ \mathcal{M} $ in Eq.~(\ref{eq:IOM}), we can identify the hierarchy of higher-rank conserving zilches, which contain high-order derivatives of electromagnetic fields.  Substituting ${\cal M} = (-1)^{n + m} p^{2n} {\cal H}^{2m + 1}$ into Eq.~(\ref{eq:Zvac1}) (apparently, it commutes with $ \mathcal{H} $), we obtain the following conserving quantities
\begin{equation}
  \label{eq:Zmn}
  C_\chi^{(m,n)} = \frac{1}{2} \int d^3 r\left (
    \bm B \cdot \nabla^{2n} \pt^{2m + 1} \bm E - \bm E \cdot \nabla^{2n}
    \pt^{2m + 1} \bm B  \right ),
\end{equation}
which were found previously by another methods (see e.~g. Refs.~\cite{Drummond1999, Drummond2006, Philbin2013}).


\section{General formalism in media}
\label{sec:gen}
The formalism of nongeometric symmetries can be generalized to Maxwell's equations in medium, where symmetry of constituent relations imposes additional constrains on the form of conservation laws. In general, this leads to the reduction of original eight-dimensional invariance algebra $A_{8}$ to smaller number of elements.
In this section, we analyze the situation when electromagnetic field propagates in time-independent dielectric medium. We show that, basically, $A_{8}$ shrinks to one of its commutative subalgebras $A_{4}$ with four basis elements. In the subsequent sections, we demonstrate that this situation is common for isotropic chiral media as far as for chiral gyrotropic crystals when light spreads along the principal symmetry direction. 
It should be mentioned that the existence of $A_{4}$ symmetry does not guarantee the conservation of optical chirality, as we discuss at the end of this section.

On a macroscopic level, the Maxwell's equations in dielectric media can be expressed in the following form
\begin{align}
\label{eq:Max1}
\vn \times \bm E = -\pt \bm B,   \qquad &
\vn \times \bm H =  \pt \bm D,   \\
\label{eq:Max2}
\vn \cdot \bm D = 0,             \qquad &
\vn \cdot \bm B = 0,          
\end{align}
which should be accompanied by constituent relations. 
In what follows, we consider the following form of  constituent relations 
\begin{equation}
\label{eq:crel}
\left( 
\begin{array}{c}
\bm D(t, \bm p) \\
\bm B(t, \bm p)
\end{array}
\right) = \mathcal{A}(\bm p)
\left( 
\begin{array}{c}
\bm E(t, \bm p) \\
\bm H(t, \bm p)
\end{array}
\right),
\end{equation}
where the momentum representation is used. The matrix $\mathcal{A}(\bm p)$ is supposed to be time-independent and determined by the properties of medium.

For symmetry analysis in medium, it is convenient to introduce Silberstein-Bateman vector $\psi(t,\bm p) = (\bm D, \bm B)^T$. In this case, the first pair of the Maxwell's equations is written as
\begin{equation}
  \label{eq:L1A}
  {\cal L}^{(\mathcal{A})} \psi = 0, \qquad
  {\cal L}^{(\mathcal{A})} = i \pt - {\cal H}^{(\mathcal{A})},
\end{equation}
where
\begin{equation}
  \label{eq:Ha}
  {\cal H}^{(\mathcal{A})} = -\sigma_2 \otimes (\hat{\bm{S}} \cdot \bm{p}) \, 
  \mathcal{A}^{-1},
\end{equation}
and the constraint on $ \psi(t, \bm p) $ imposed by the second pair of Maxwell's equations is the same as  in Eq.~(\ref{eq:L3}).

For a common situation when constituent relations do not mix up electric and magnetic fields, Eq.~(\ref{eq:Ha}) is reduced to the following expression
\begin{equation}
  \label{eq:H1}
  {\cal H}^{(\mathcal{A})} = 
  \begin{pmatrix}
  0 & i (\hat{\bm{S}} \cdot \bm{p}) \hat{\mu}^{-1}(\bm{p}) \\
  -i(\hat{\bm{S}} \cdot \bm{p}) \hat{\ee}^{-1}(\bm{p}) & 0   
  \end{pmatrix},
\end{equation}
where $\hat{\ee}^{-1}(\bm{p})$ and $\hat{\mu}^{-1}(\bm{p})$ are inverse permittivity and permeability
tensors.

\subsection{Nongeometric symmetries}
In general, it may be a tedious problem to find symmetry transformations for the Maxwell's equations in medium.  However, the task is alleviated along the directions where transverse electromagnetic waves can propagate. Mathematically, it corresponds to $\mathcal{A}(\bm p)$ being commutative with $\mathcal{L}$ in Eq.~(\ref{eq:L3}) along such directions.  In this case, to identify possible symmetry transformations  we apply a transformation $\bar{\psi} = \mathcal{N}^{-1} \psi$ to the basis, where $\mathcal{H}^{(\mathcal{A})}$ and $\mathcal{L}$ are both diagonal
\begin{eqnarray}
\label{eq:rot}
\bar{{\cal H}}^{({\cal A})}  &=&   
\mathcal{N}^{-1} {\cal H}^{({\cal A})} \mathcal{N} = \diag(\omega_1,\omega_2,0,\omega_4,\omega_5,0), \\
\label{eq:rot11}
\bar{{\cal L}}  &=&  \mathcal{N}^{-1} {\cal L} \mathcal{N}
= \diag(0,0,p^2,0,0,p^2),
\end{eqnarray}
where parameters $\omega_{i} = \omega_{i}(\bm{p})$ are real-valued in the absence of dissipation. Note that, since in the original basis $\mathcal{H}^{(\mathcal{A})}$ is not necessary a Hermitian matrix, the transformation $\mathcal{N}$ may be non-unitary. 

For the diagonal operators in Eqs.~(\ref{eq:rot}, \ref{eq:rot11}), it is easy to find the invariance algebra. The number of basis elements in the invariance algebra depends on the symmetry relations between $\omega_i$ (see Appendix~\ref{app:IA} for mathematical aspects of the derivation). In the absence of any degeneracies between $\omega_i$, all symmetry transformations in the $\bar{\psi}$-basis are given by four diagonal operators. The basis in this four-dimensional linear space can be chosen as 
\begin{eqnarray}
  \label{eq:QBrot}
  \bar{\mathcal{Q}}_{a} &=& \diag(1,1,0,-1,-1,0), \\ \label{eq:Qb}
  \bar{\mathcal{Q}}_{b} &=& \diag(-1,1,0,-1,1,0), \\
  \bar{\mathcal{Q}}_{c} &=& \diag(1,1,0,1,1,0),   \\ \label{eq:QBrot2}
  \bar{\mathcal{Q}}_{d} &=& \diag(-1,1,0,1,-1,0),
\end{eqnarray}
which can be conveniently expressed by the Kronecker product of $\{\sigma_0, 
\sigma_3\}$ and $\{\hat I, \hat{\Gamma}\}$ where $\hat{\Gamma} = \diag(-1,1,0)$,
that form the Klein four-group isomorphic to the
direct sum $\mathbb{Z}_{2} \oplus \mathbb{Z}_{2}$ \cite{Robinson2012}. 
The form of these operators in the original basis $\psi$ is reached 
by the inverse transformation $\psi  = {\cal N} \bar{\psi}$. 

From Eqs.~(\ref{eq:QBrot})--(\ref{eq:QBrot2}), we make a conclusion that radiation field in media is invariant at least under the four commutative symmetry transformations, which we will refer to as \emph{$ A_4 $ symmetries}. This group has two trivial elements that correspond to the identity and
$ i\partial_t \equiv \mathcal{H}^{\mathcal{A}} $. The latter simply states that in time-independent medium  time derivative of the solution is again the solution. The physical meaning of the other two elements is determined by the constituent relations encoded into the transformation $ \mathcal{N} $.
Let us note that $ A_4 $ is a minimal symmetry. In the case of additional degeneracies between different $\omega_{i}$ in Eq.~(\ref{eq:rot}), $A_{4}$ becomes a subalgebra of a larger invariance algebra. 

For illustration, we consider free electromagnetic field.  
In vacuum, the diagonal form of $\mathcal{H}$ in Eq.~(\ref{eq:Hamilt}) is reached by unitary transformation $ \psi = \mathcal{U} \bar \psi $, where $\mathcal{U} = U_{2} \otimes \hat{U}_{\Lambda}$ combines the transformation to the helicity basis
\begin{equation}
  \label{eq:UL}
  \hat{U}_{\Lambda} =
  \left(
    \begin{array}{ccc}
      -\dfrac{p_1p_3 + i p_2p}{\sqrt{2}pp_\perp} & \dfrac{p_1p_3 - i p_2p}{\sqrt{2}pp_\perp} & \dfrac{p_1}{p} \\
      -\dfrac{p_2p_3 - i p_1p}{\sqrt{2}pp_\perp} & \dfrac{p_2p_3 + i p_1p}{\sqrt{2}pp_\perp} & \dfrac{p_2}{p} \\
      \dfrac{p_\perp}{\sqrt{2}p}                & - \dfrac{p_\perp}{\sqrt{2}p}              & \dfrac{p_3}{p}
    \end{array}
  \right),
\end{equation}
where $ (\hat{\bm S} \cdot \bm p) $ is diagonal, with $SU(2)$ rotation in the pseudospace of $\bm{D}$ and $\bm{B}$ vectors
\begin{equation}
  \label{eq:U2}
  U_2 = \frac{1}{\sqrt{2}}\left(\sigma_0 - i\sigma_1\right).
\end{equation}
Comparing the resulting diagonal form of $ \mathcal{H} $
\begin{equation}
  \label{eq:rot1}
  \bar{{\cal H}} = \mathcal{U}^{\dag} \mathcal{H} \mathcal{U} = p\sigma_3 \otimes \hat{\Gamma} = \diag(-p, p, 0, p, -p , 0),
\end{equation}
with Eq.~(\ref{eq:rot}), we find that $ \bar{\mathcal{H}} $ has two degeneracies between $ \omega_{i} $, namely, $\omega_1 = \omega_5 = -p$ and $\omega_2 = \omega_4 = p$.  The existence of these degeneracies, according to (\ref{eq:[QH]}), means that the symmetry transformation $ \bar{\mathcal{Q}}_A $ in (\ref{eq:Qa}) contains in total eight free parameters: $ q_{11} $, $ q_{22} $, $ q_{44} $, $ q_{55} $, $ q_{15} $, $ q_{24} $, $ q_{51} $, and $ q_{42} $. 
In the original basis $ \psi = \mathcal{U}^{-1} \bar{\psi} $, it gives rise to the  eight-dimensional algebra $A_8$ in Eqs.~(\ref{eq:QA})--(\ref{eq:Q8}). Under the inverse transformation, four diagonal operators in Eqs.~(\ref{eq:QBrot})--(\ref{eq:QBrot2}) transform into: 
$\bar{\mathcal{Q}}_{a} \to \mathcal{Q}_{2}$, 
$\bar{\mathcal{Q}}_{b} \to \mathcal{Q}_{5}$, 
$\bar{\mathcal{Q}}_{c} \to \mathcal{Q}_{7}$, and
$\bar{\mathcal{Q}}_{d} \to \mathcal{Q}_{8}$.

\subsection{Optical chirality}
The generalization of the conservation laws in Eq.~(\ref{eq:IOM})  
in media is straightforward. We define conserving quantities
\begin{equation}
\label{eq:IOM3}
\langle \mathcal{Q}_{A} \rangle = 
\frac{1}{2}\int d^{3} p \, \psi^{\dag}(t,\bm{p}) \, \rho(\bm p)
\mathcal{Q}_{A}(\bm{p}) \,\psi(t,\bm{p}),
\end{equation} 
where $ \mathcal{Q}_{A} $ is one of the symmetry transformations.
The scalar product is modified by $ \rho =  (\mathcal{N}^{-1})^{\dag} \mathcal{N}^{-1}$  for non-unitary $ \mathcal{N} $. For any Hermitian matrix $ \mathcal{H}^{\mathcal{A}} $, the property
$ \rho = 1 $ is restored.

In general, the existence of the symmetries defined in 
Eqs.~(\ref{eq:QBrot})--(\ref{eq:QBrot2}) does not automatically ensure  the 
conservation law for optical chirality. It can be introduced if the 
invariance algebra contains an element that yields a conserving
pseudoscalar $C_{\chi}$, which is  simultaneously 
even under time-reversal and odd under spatial inversion symmetries.

The existence of $ C_{\chi} $ is justified in the medium that has duality symmetry \cite{Berry2009, Fernandez2013}. In this case, the  duality transformation
$ \mathcal{Q}_{\mathrm{dual}} = i \rho^{-1} (\sigma_2 \otimes \hat{I})$ 
is one the symmetries, and $ C_{\chi} $ can be introduced as a 
conservation of $ \mathcal{Q}_{A} = \mathcal{Q}_{\mathrm{dual}} \pt $ in Eq.~(\ref{eq:IOM3})
that eventually leads to the following form of optical chirality 
\begin{equation}
\label{eq:Cchi1}
C_{\chi}^{\mathrm{(dual)}} = \frac{1}{2} \int d^{3} p \, 
\left( \bm{B}^{*} \cdot \pt \bm{D} - \bm{D}^{*} \cdot \pt \bm{B} \right).
\end{equation}
The general form of the constituent relations in media, which preserves duality transformation was obtained in Ref.~\cite{Fernandez2013}. It was demonstrated that necessary and sufficient condition for the system to remain self-dual is commutativity of the duality transformation generator and the matrix of constituent relations.

Unlike free electromagnetic field, where conservation of  $\mathcal{Q}_{\mathrm{dual}}\pt$
is equivalent to the helicity conservation (see Eqs.~(\ref{eq:Zvac}, \ref{eq:Zvac1})),
in media it is possible that the helicity is conserved even if the system does not have 
dual symmetry. In this case, we define $ C_{\chi} $ as a conservation of $(\bm{\mathcal{S}} \cdot \bm p) $ operator in Eq.~(\ref{eq:IOM3}). A general criterion for the helicity conservation is commutativity of $(\bm{\mathcal{S}} \cdot \bm p) $ operator with the matrix of constituent relations $ \mathcal{A} $.

Note that there is no ambiguity in the definition of optical
chirality, if the system is invariant under both the duality symmetry and the helicity transformations.
In this case, these two elements belong the same set of transformations in Eqs.~(\ref{eq:QBrot})--(\ref{eq:QBrot2}), which means that the product  $\mathcal{Q}_{\mathrm{dual}} \pt $ is a linear combination of other symmetry elements of $ A_4 $ that include $ (\bm{\mathcal{S}} \cdot \bm p) $.


\section{Optical chirality in media}
\label{sec:iso}
In dielectric media, optical activity is a usual manifestation of microscopic structural chirality.  This effect, in general, is related to noncentrosymmetry and shared by both chiral and achiral materials \cite{Claborn2008}.  On a macroscopic level, optical activity can be described by proper constituent relations. Historically, constituent relations describing natural optical activity were first developed in Born-Drude-Fedorov (BDF) form
\cite{Condon1937, Born1972, Fedorov1959, Fedorov1976}
\begin{eqnarray}
\label{eq:Fedorov}
\bm{D} &=& \hat{\varepsilon} \varepsilon_{0} \left(\bm{E} + \hat{\alpha} \bm{\nabla}
\times \bm{E}\right), \\
\label{eq:Fed1}
\bm{B} &=& \hat{\mu}         \mu_{0}      \left(\bm{H} + \hat{\alpha}^{T} \bm{\nabla}
\times \bm{H}\right),
\end{eqnarray}
where $\hat{\ee}$ and $\hat{\mu}$ are the electric permittivity and the
magnetic permeability tensors and $\hat{\alpha}$ is the gyration tensor ($\hat \alpha^{T}$ means transposed $ \hat \alpha $).

Another form of constituent relations that features optical rotation, which
we will refer to as chiral magnetoelectric (CME) constituent relations,
comes from the  general relativity covariance principle \cite{Fedorov1976, Post1962}
and can be written as follows
\begin{eqnarray}
\label{eq:Post}
\bm{D} &=& \hat{\varepsilon} \varepsilon_{0} \bm{E} + i \hat{\vk} \bm{H}, \\
\label{eq:Pos1}
\bm{B} &=& \hat{\mu}         \mu_{0}         \bm{H} - i \hat{\vk}^{T} \bm{E},
\end{eqnarray}
where $\hat{\vk}$ is the magnetoelectric tensor \cite{Rado1962}.  We note that these relations are usually formulated in the frequency domain for time-harmonic electromagnetic fields with the frequency factor $ \omega $ being included in $ \hat{\vk} $.  In what follows, when we use CME relations, we imply that the formalism of complex time-harmonic fields is used, which time dependencies are given by $ \exp(-i\omega t) $.

CME form of constituent relations is frequently used in chiral metamaterials \cite{Jaggard1979, Tomita2014} and in crystals with gyrotropic birefringence \cite{Hornreich1968, Hornreich1968a}. Mutual relation between BDF and CME equations has been studied by several authors \cite{Lekner1996, Firsanov2007, Cho2015}.

\subsection{Isotropic chiral media}
Let us first consider optical chirality in isotropic chiral media characterized by 
constituent relations of either BDF or CME type.  Both types share a number of
common features, therefore,  we carry out the discussion in parallel.  
To avoid redundant complications, 
we use the units where $\ee\ep = \mu\mu_0 =1$ and restore SI units whenever it is necessary.

The matrix form of constituent relations in the momentum space 
is given by Eq.~(\ref{eq:crel}) with 
\begin{equation}
\mathcal{A}(\bm p) = 
\begin{dcases*}
\sigma_0 \otimes (\hat{I} + \alpha (\hat{\bm{S}} \cdot \bm{p})), & BDF, \\
(\sigma_0 - \vk \sigma_2) \otimes \hat I, & CME,
\end{dcases*}
\end{equation}
where the upper (lower) line is for constituent relations in 
Eqs.~(\ref{eq:Fedorov}, \ref{eq:Fed1})
(Eqs.~(\ref{eq:Post}, \ref{eq:Pos1})),
that provides Hermitian matrix
\begin{equation}
\mathcal{H}^{(\mathcal{A})} = 
\begin{dcases*}
-\sigma_2 \otimes \dfrac{(\hat{\bm S} \cdot \bm p) - \alpha (\hat{\bm S} \cdot \bm p)^{2}}{1 - \alpha^2 p^2}, & BDF, \\
-(\sigma_2 + \vk \sigma_0) \otimes \dfrac{(\hat{\bm S} \cdot \bm p)}{1 - \vk^2}, & CME,
\end{dcases*}
\end{equation}
which is invariant under both duality and helicity transformations.

To find the complete set of the nongeometric symmetries, we diagonalize $\mathcal{H}^{(\mathcal{A})}$
by applying the same unitary transformation as in vacumm, ${\cal U} = U_2 \otimes
\hat{U}_\Lambda$, where $\hat{U}_{\Lambda}$ and $U_2$ 
are defined in Eqs.~(\ref{eq:UL}) and (\ref{eq:U2}), respectively, which leads to the diagonal form
\begin{equation}
  \bar{\mathcal{H}}^{(\mathcal{A})} 
  =  \diag(-p_{\pm}, p_{-}, 0, p_{+}, -p_{\mp}, 0),
\end{equation}
with upper (lower) sign for BDF (CME) constituent relation, and
$ p_{\pm} = p (1 \mp \alpha p)^{-1} $ ($ p_{\pm} = p (1 \mp \vk)^{-1} $) for BDF (CME).

In chiral media, symmetry breaking between left and right polarized states removes
the degeneracy between the eigenvalues of $ \bar{\mathcal{H}}^{(\mathcal{A})}$,
and according to Eqs.~(\ref{eq:rot}) and (\ref{eq:Qa}), the set of nongeometric 
symmetries is reduced to four elements with the following basis 
\begin{align}
  \label{eq:Qbdf}
  \mathcal{Q}_{2} = i \sigma_{2} \otimes \hat{I},               \qquad &
  \mathcal{Q}_{5} =   \sigma_{0} \otimes (\hat{\bm{S}} \cdot \tilde{\bm{p}}),         \\
  \label{eq:Qbdf1}
  \mathcal{Q}_{7} =   \sigma_{0} \otimes \hat{I},               \qquad &
  \mathcal{Q}_{8} = i \sigma_{2} \otimes (\hat{\bm{S}} \cdot \tilde{\bm{p}}),
\end{align}
which includes both duality ($ \mathcal{Q}_{2} $) and helicity ($ \mathcal{Q}_{5} $) transformations.
Therefore, in isotropic chiral media, similar to the case of free electromagnetic field,  
we can say that duality symmetry is related to the helicity conservation.

Since duality symmetry is preserved, Lipkin's zilch is directly obtained from Eq.~(\ref{eq:Cchi1}) that  in the $ \bm{r} $-space is written as
\begin{equation}
  \label{eq:Cbrel}
  C_{\chi}^{(\mathrm{iso})} = \frac{1}{2} \int d^3 r \left ( \bm B^{*} \cdot
    \pt \bm D - \bm D^{*} \cdot \pt \bm B\right ).
\end{equation}
Here, and in Eqs.~(\ref{eq:Cb1})--(\ref{eq:Fsrc}), we hold the following convention.  For BDF constituent relations one has to remove complex conjugation for the fields in $ \bm{r} $-space. In contrast, for CME relations all the fields are supposed to be time-harmonic complex fields, and $ \partial_{t} $ should be replaced by $ -i\omega $ in final expressions. The transformation to SI units in Eq.~(\ref{eq:Cbrel}) is rendered by the substitutions 
\begin{align*}
\bm{D} &\to \bm{D} (\ee\ep)^{-1},                         &
\bm{B} &\to \bm{B} (\ee\ep\mu\mu_0)^{-\frac{1}{2}},              \\
\bm{H} &\to \bm{H} 
\left( \dfrac{\mu\mu_0}{\ee\ep}\right)^{\frac{1}{2}},     &
t &\to t (\ee\ep\mu\mu_0)^{-\frac{1}{2}},
\end{align*}
supplemented by $C_{\chi}^{(\mathrm{iso})} \to \ee\ep C_{\chi}^{(\mathrm{iso})}$.

We emphasize that in infinite homogenious medium optical chirality can be expressed in 
several equivalent forms. For example, instead of the operator $ \mathcal{Q}_2 \pt $ 
that gives the conservation law in Eq.~(\ref{eq:Cbrel}), we can consider another symmetry operation
$ \mathcal{Q}_2 \mathcal{A}^{-1} \pt $, which  leads to 
\begin{equation}
  \label{eq:Cb1}
  C_{\chi}^{(\mathrm{iso})} 
  =  \frac{1}{2} \int d^3 r \left ( \bm B^{*} \cdot
    \pt \bm E - \bm D^{*} \cdot \pt \bm H \right ).
\end{equation}
However, in the realistic case, we should also care about the conservation of chirality
flow across the boundaries separating different media, which removes this ambiguity.

The situation with several forms of optical chirality is not new. Similar situation happens with the energy density in chiral materials. In the absence of boundaries, energy density can be also expressed in several equivalent forms. However, only one form  guarantees proper energy balance across the boundary
between two chiral media \cite{Fedorov1976}. It was demonstrated by Fedorov \cite{Fedorov1976}
that physical form of energy density depends on the choice of constituent relations as follows
\begin{equation}
\mathfrak{E} = \frac{1}{2}
\begin{dcases*}
(\varepsilon \varepsilon_0)^{-1} \bm{D} \cdot \bm{D} 
+ (\mu \mu_0)^{-1} \bm{B} \cdot  \bm{B}, & BDF, \\
\bm{E}^{*} \cdot \bm{D} + \bm{H}^{*} \cdot \bm{B}, & CME.
\end{dcases*}
\end{equation}
We anticipate that similar to energy density, optical chirality for BDF and CME 
constituent relations should be taken in different forms.

The situation becomes more transparent in the spatially nonuniform space
where $\ee(\bm r)$, $\mu(\bm r)$, $\alpha(\bm r)$, and $\vk(\bm r)$ 
depend on the local position.
Looking for the proper form of zilch density in the real space, we have settled on the following choice
\begin{equation}
  \label{eq:rho}
  \rho_{\chi} = \frac{1}{2}
  \begin{dcases*}
    \bm B \cdot \pt \bm D - \bm D \cdot \pt \bm B, & BDF \\
    \ee\ep \, \bm B^{*} \cdot \partial_t \bm E - \mu \mu_0 \, 
    \bm D^{*} \cdot \partial_t \bm H, & CME,
  \end{dcases*}
\end{equation} 
which corresponds to $C_{\chi}^{(\mathrm{iso})}$ given by Eq.~(\ref{eq:Cbrel}) 
(Eq.~(\ref{eq:Cb1})) for BDF (CME) medium.

In the nonuniform space, the conservation law for $\rho_{\chi}$ is violated by the 
source term on the right hand side of the continuity equation 
\begin{equation}
  \label{eq:drho}
  \pt \rho_{\chi} + \vn \cdot \bm J_{\chi} 
  = F(t,\bm r). 
\end{equation}
However, both expressions for zilch density in Eq.~(\ref{eq:rho}) 
are related to the same zilch flow
\begin{equation}
  \bm J_{\chi} = \frac{\ep   \ee   }{2} \, \bm E^{*} \times \partial_{t} \bm E +
                 \frac{\mu_0 \mu   }{2} \, \bm H^{*} \times \partial_{t} \bm H,
\end{equation}
and the source term
\begin{equation}
  \label{eq:Fsrc}
  F(t, \bm r) = 
    \frac{\ep  }{2} \, \vn \varepsilon \cdot \bm E^{*} \times \partial_{t} \bm E + 
    \frac{\mu_0}{2} \, \vn \mu         \cdot \bm H^{*} \times \partial_{t} \bm H.
\end{equation}

The source term contains only gradients of $\ee$ and $\mu$. 
In this regards, we mention Ref.~\cite{Bialynicki1996} where it has been demonstrated 
that in isotropic time-independent media the mixing of helicity occurs only in a presence of the space-dependent ``resistance'' proportional to $\sqrt{\mu(\bm r)/\varepsilon(\bm r)}$.
Similar to Ref.~\cite{Bialynicki1996},  
the absence of the gradients of gyrotropic constants in $ F(t, \bm r) $
justifies continuity of $ \bm J_{\chi} $ between two chiral media if 
$\sqrt{\mu/\varepsilon}$ remains the same across the boundary 
\footnote{%
If the relation $ \mu(\bm r)/\varepsilon(\bm r) = \mathrm{Const}  $ holds in medium
with constituent relations in Eqs.~(\ref{eq:Fedorov}--\ref{eq:Pos1}), this medium is self-dual according to 
Ref.~\cite{Fernandez2013}.  In this case, we can also rewrite Eq.~(\ref{eq:drho}) in the form of conservation law for redefined $ \rho_\chi $ and $ \bm J_\chi $
}.

We emphasize that the absence of $ \bm \nabla \alpha $ or $ \bm \nabla \vk  $ 
in the source term takes place only for the 
form of $ \rho_{\chi} $ in Eq.~(\ref{eq:rho}).



\begin{table*}
\caption{Parameters and point groups in crystals with gyrotropic birefringence}
\label{tab:bigyr}
\begin{ruledtabular}
\begin{tabular}{ll}
Tensors & Point groups \\
\hline
\multirow{3}{*}{
$\hat{\ee} = \left(
\begin{array}{ccc}
 \ee_\perp  & -ia &  0 \\
 ia &  \ee_\perp  &  0 \\
 0 &  0 & \ee_\parallel 
\end{array}
\right), \qquad
\hat{\mu} = \left(
\begin{array}{ccc}
 \mu_\perp  & -ib &  0 \\
 ib &  \mu_\perp  &  0 \\
 0 &  0 &  \mu_\parallel 
\end{array}
\right)$ 
               } 
& Tetragonal: $C_{4}$ ($4$), $C_{4h}$ ($4/m$), $S_{4}$ ($\bar{4}$) \\
& Trigonal: $C_{3}$ ($3$), $S_{6}$ ($6$)                           \\
& Hexagonal: $C_{6}$ ($6$), $C_{3h}$ ($\bar{6}$), $C_{6h}$ ($6/m$) \\
$\hat{\ee} \hat{\mu} = \hat{\mu}\hat{\ee}, \quad 
\bm{a} \parallel \bm {b} \parallel \hat{z}$ & \\
\end{tabular}
\end{ruledtabular}
\end{table*}


\subsection{Optical chirality in crystals}
\label{sec:cryst}
In crystals, nonequivalent directions have different symmetries that are encoded in the structure of material tensors. Therefore, our general formalism should be applied with respect to certain crystalline directions.
We note that, in this section, under the duality and helicity symmetries, we mean the symmetry transformations with respect to the principal axis. This resembles the situation with the forward and backward scattering symmetry theorems, where the explicit form of the Mueller matrix for the light scattering shows features  that are related to the crystal symmetry of the dielectric scatter \cite{Hu1987}.

Fundamentally, anisotropic gyrotropic media are split into two different groups,
namely, crystals with gyrotropic birefringence and media with natural 
optical activity, which show different behavior with respect to mirror 
reflections \cite{Fedorov1976}. 

Let us first briefly focus on crystals with gyrotropic birefringence.
These materials are characterized by the following constituent relations 
\cite{Hornreich1968, Hornreich1968a}
\begin{eqnarray}
\label{eq:Big1}
\bm D = \left(\hat{\ee}_s + i\bm a \times \right) \bm E = \hat{\ee} \bm E, \\
\label{eq:Big2}
\bm B = \left(\hat{\mu}_s + i\bm b \times \right) \bm H = \hat{\mu} \bm H,
\end{eqnarray}
with $\bm a$ and $\bm b$ being the gyrotropic vectors, 
and $\hat{\ee}_s$  ($\hat{\mu}_s$) stands for the diagonal part 
of $\hat{\ee}$ ($\hat{\mu}$).  We consider the case when 
$\bm a$ and $\bm b$ are parallel to the high symmetry direction taken as
$z$-axis. The explicit form of $\hat{\ee}$ and  $\hat{\mu}$ for some point groups is given in Table~\ref{tab:bigyr}. As it is well known, this kind of gyrotropy is prohibited in cubic crystals \cite{Fedorov1976}.

The constituent relations in Eqs.~(\ref{eq:Big1}, \ref{eq:Big2}) preserve the duality symmetry
\cite{Fernandez2013}. Apparently, along the $ z $-axis the helicity operator $ \hat S_z $
is also a symmetry transformation, since it commutes with the constituent relations. 
Taking into account broken inversion symmetry, we conclude that similar to the isotropic
chiral media, the set of nongeometric symmetries in birefringent crystals for $ \bm p \parallel \hat z $ is four-dimensional. It contains identity, $ i\partial_t $, $ \hat S_z $, and duality $\mathcal{Q}_{\mathrm{dual}}$. The latter guarantees conservation of chirality 
in the form of  Eq.~(\ref{eq:Cchi1}).

Maxwell's equations in crystals with natural optical activity are 
given by Eqs.~(\ref{eq:L1A})--(\ref{eq:H1}) with the following 
permittivity and permeability tensors
\begin{align}
\label{eq:con1}
\hat{\ee}(\bm p) = & \hat{\ee}_{s} \left(\hat{I} 
                     + \hat{\alpha}(\hat{\bm{S}} \cdot \bm{p})\right),
\\
\label{eq:con2}
\hat{\mu}(\bm p) = & \hat{\mu}_{s} \left(\hat{I} 
                     + \hat{\alpha}^{T}(\hat{\bm{S}} \cdot \bm{p})\right).
\end{align}
Henceforth, we point $\bm{p}$ along the symmetry axis, $\bm p = p \hat{z}$.
Different forms of the gyration tensor for point groups in cubic, tetragonal, and hexagonal crystal families
are listed in Table~\ref{tab:alpha}. In these crystal families, $\hat{\varepsilon}_{s}$ and $\hat{\mu}_{s}$ 
are given by the diagonal parts of $\hat{\ee}$ and $\mu$ in Table~\ref{tab:bigyr}. Apparently, the cubic case  is identical to isotropic media.

All point groups in Table~\ref{tab:alpha} are noncentrosymmetric and
break down into chiral and achiral parts. The former are represented by the point 
groups of eleven enantiomorphic pairs of chiral space groups, namely,
$T$, $O$, $C_{n}$ and $D_{n}$ ($n \ge 3$) \cite{Glazer2013}, while the 
latter are given by $S_{4}$, $D_{2d}$, and $C_{nv}$.
For  a review of natural optical activity in achiral materials see,
for instance, \cite{Fedorov1976, Claborn2008}.

\begin{figure}
  \centerline{\includegraphics[scale=.3]{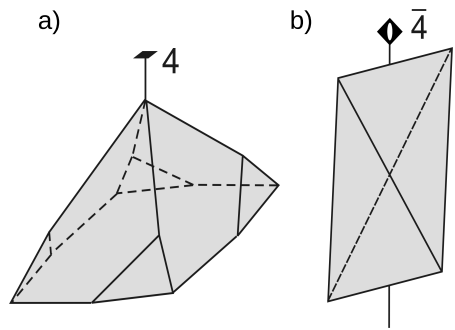}}
  \caption{Crystals with (a) $C_{4}$  and  (b) $S_{4}$ point groups.}
  \label{fig:1}
\end{figure}

Let us consider symmetry transformations in crystals with natural optical activity. According to the conditions for dual systems (see Eq.~(11) in Ref.~\cite{Fernandez2013}), we find that the constituent relations in Eqs.~(\ref{eq:con1}, \ref{eq:con2}) preserve the duality symmetry only if $ \hat \alpha = \hat \alpha^{T} $. This means that the duality symmetry is broken in $ C_n $ and $ C_{nv} $ ($n \ge 3$) point groups, see Table~\ref{tab:alpha}.

At the same time, $ \hat\alpha $ commutes with the helicity transformation $ \hat S_z $
in $ C_n $ and $ C_{nv} $ ($n \ge 3$), while in $ D_{2d} $ and $ S_4 $ commutativity between
$ \hat \alpha $ and $ \hat S_z $ does not hold. The conservation of helicity in $ C_n $ and $ C_{nv} $ is supported by the scattering theorem, which states that electromagnetic forward scattering in linear systems with the discrete rotational symmetry ($ n \ge 3 $) can be only helicity preserving when light spreads along the principal axis \cite{Fernandez2013a}.

The examples above demonstrate that in crystals with natural optical activity it is possible
that either duality or helicity operator is the symmetry transformation but not both of them
at the same time. We consider these two cases in more detail in the following sections.

\subsubsection{Conservation of helicity with broken duality}
Let us consider nongeometric symmetry transformations in $C_{n}$ point groups
($ n \ge 3 $). The  gyration tensor $\hat{\alpha}$ is given in Table~\ref{tab:alpha}.
For illustration, a chiral crystal belonging to the $C_4$ point group is shown in 
Fig.~\ref{fig:1}~(a).

The diagonal form of $\mathcal{H}^{(\mathcal{A})}$ in Eq.~(\ref{eq:H1}) is
brought forth by the unitary transformation
\begin{equation}
\label{eq:UC}
\psi(t, \bm{p}) = \mathcal{U}_{C} \bar{\psi}(t, \bm{p})
\end{equation}
(see Appendix~\ref{app:C4} for explicit form of $\mathcal{U}_{C}$) that brings about 
\begin{equation}
\label{eq:Ha1}
\bar{\mathcal{H}}^{\mathcal{A}} = 
\diag\left(p_{+}, -p_{-}, 0, -p_{+}, p_{-}, 0 \right),
\end{equation}
with $ p_{\pm} = p[(1 \mp \alpha_{0} p)^{2} + \alpha_{2}^{2}p^{2}]^{-1/2}$.

The form of Eq.~(\ref{eq:Ha1})  suggests that in $C_{n}$ point groups 
the invariance algebra for the light spread along the high symmetry direction 
is four-dimensional $ A_{4} $. In the transformed frame, the basis elements are given in Eqs.~(\ref{eq:QBrot})--(\ref{eq:QBrot2}). However, in achiral point groups $C_{nv}$, we have additional constraint $\alpha_0 = 0$ (see Table~\ref{tab:alpha}), which restores the symmetry between left and right polarized states, $p_{+} = p_{-}$, and the resulting invariance algebra becomes eight-dimensional, 
in agreement with Eq.~(\ref{eq:[QH]}).

In order to define optical chirality, we construct the following operator
in the transformed frame
\begin{equation}
\bar{\mathcal{Q}}_{\chi}(\bm{p}) =
\frac{p}{2p_{-}}\left(\bar{\mathcal{Q}}_{a} + \bar{\mathcal{Q}}_{d}\right)
+ \frac{p}{2p_{+}}\left(\bar{\mathcal{Q}}_{a} - \bar{\mathcal{Q}}_{d}\right),
\end{equation}
which in the original basis is written as
\begin{equation}
\label{eq:Q11}
\mathcal{Q}_{\chi}(\bm{p}) = 
\begin{pmatrix}
0 & -i\hat{I} - i \hat{\alpha} (\hat{\bm S} \cdot \bm p) \\
i\hat{I} + i \hat{\alpha}^{T} (\hat{\bm S} \cdot \bm p)
\end{pmatrix}.
\end{equation}

We can use this operator to define optical chirality. 
Using $ \mathcal{Q}_A  = i \mathcal{Q}_{\chi} \pt$ in Eq.~(\ref{eq:IOM3}), 
optical chirality in point groups $C_{n}$, $C_{nv}$ is obtained as
\begin{multline}
C_{\chi}^{(c)} = \frac{1}{2} \int d^{3} p
\left( 
\bm{D}^{*} \cdot \pt \bm{B} - \bm{B}^{*} \cdot \pt \bm{D}
\right)\\
+ \frac{i}{2} \int d^{3} p 
\left( \bm{D}^{*} \cdot \hat{\alpha} \bm p \times \pt \bm{B} 
- \bm{B}^{*} \cdot \hat{\alpha}^{T} \bm p \times \pt \bm{D}\right).
\end{multline}
By using the constituent relations in Eqs.~(\ref{eq:con1}, \ref{eq:con2}), 
this equation can be rewritten in a compact form
\begin{equation}
\label{eq:100}
C_{\chi}^{(c)} = -\frac{i}{2} \int d^{3} p
\left( \bm{D}^{*} \cdot \ee_{s}^{-1}\bm{p} \times \bm{D} + 
\bm{B}^{*} \cdot \mu_{s}^{-1} \bm{p} \times \bm{B} \right),
\end{equation}
which is nothing but the conservation law for the helicity operator 
$ (\bm{\mathcal{S}} \cdot \tilde{\bm p}) $, see Eq.~(\ref{eq:A9}) in Appendix~\ref{app:C4}.

\subsubsection{Duality symmetry without helicity transformation}
The gyration tensor in point groups $S_{4}$ and $D_{2d}$ is 
given in Table~\ref{tab:alpha}. We take $\hat{\alpha}$ in the form 
that corresponds to $S_{4}$ group. The case of $D_{2d}$ group is obtained by setting
$\alpha_{0} = 0$. An example of achiral crystal with
$S_{4}$ point group is demonstrated in Fig.~\ref{fig:1}~(b).

The unitary transformation that diagonalizes $\mathcal{H}^{(\mathcal{A})}$ in 
$S_{4}$ and $D_{2d}$ is defined as
\begin{equation}
\label{eq:US}
\psi(t, \bm{p}) =
\mathcal{U}_{S} \bar{\psi}(t, \bm{p}), 
\end{equation}
where $\mathcal{U}_{S}$ is specified in Appendix~\ref{app:S4}, which leads to
the following dialgonal form in Eq.~(\ref{eq:rot}) 
\begin{equation}
\label{eq:Hac}
\bar{\mathcal{H}}^{(\mathcal{A})} = \diag\left(-p_{1}, p_{1}, 0, p_{1}, -p_{1}, 0 \right),
\end{equation}
where $p_{1} = p[1 + (\alpha_{0}^{2} + \alpha_{2}^{2})p^{2}]^{-1/2}$.

Equation~(\ref{eq:Hac}) shows that unlike $C_{n}$ point groups,
the symmetry properties of $ \mathcal{H}^{\mathcal{A}} $ in $S_{4}$ and $D_{2d}$, for $\bm p$ along the principal axis,  are similar to the case of $C_{nv}$ point groups. The invariance algebra of the symmetry transformations in $S_{4}$ and $D_{2d}$ is eight-dimensional, since the symmetry between left and right polarized states remains unbroken, which is related to the absence of optical rotation along the symmetry direction in achiral crystals \cite{Claborn2008}.

Since in $S_{4}$ and $D_{2d}$ we have $ \hat \alpha = \hat \alpha^T $ for $ \bm{p} \parallel \hat z $, the constituent relations preserve the  duality symmetry, which means that $ \mathcal Q_{\mathrm{dual}} $ enters to the invariance algebra. This means that optical chirality in $S_{4}$ and $D_{2d}$ groups is given by Eq.~(\ref{eq:Cchi1}).

We emphasize that the helicity operator $(\bm{\mathcal{S}} \cdot \tilde{\bm{p}})$ does not belong to the symmetry transformations in $S_{4}$ and $D_{2d}$ point groups even for $ \bm{p} \parallel \hat z $.
\footnote{%
	Here, under the helicity we mean the projection of the spin onto the direction of propagation,
	$(\bm{\mathcal{S}} \cdot \tilde{\bm{p}})$. We note that to discuss the physical helicity expressed through the difference in population of left and right polarized photons, one has to construct a photon wave function in chiral crystals with $S_{4}$ and $D_{2d}$ groups, which is, however, beyond the scope of our symmetry analysis
}.  
Instead, the role of the 
helicity operator is played by $ \mathcal{Q}_{\mathrm{h}} = -p^{-1}\mathcal{Q}_{\mathrm{dual}} \partial_t$, whose explicit form is given by
\begin{equation}
\mathcal{Q}_{\mathrm{h}} = 
\begin{pmatrix}
(\bm{\mathcal{S}} \cdot \tilde{\bm{p}}) \hat{\varepsilon}^{-1}(\bm p) & 0 \\
0 & (\bm{\mathcal{S}} \cdot \tilde{\bm{p}}) \hat{\mu}^{-1}(\bm p)
\end{pmatrix},
\end{equation}
where we used the identity $ i\partial_t \equiv \mathcal{H}^{(\mathcal{A})} $ together  with 
Eq.~(\ref{eq:H1}).


\begin{table*}
	\caption{Gyration tensor in different point groups}
	\label{tab:alpha}
	\begin{ruledtabular}
		\begin{tabular}{ll|ll}
			Gyration tensor $\hat{\alpha}$ & Point groups & 
			Gyration tensor $\hat{\alpha}$ & Point groups \\
			\hline
			$
			\left(
			\begin{array}{ccc}
			\alpha_0 & \alpha_2 & 0 \\
			\alpha_2 &-\alpha_0 & 0 \\
			0     &    0     & 0
			\end{array}
			\right)
			$              
			&
			Tetragonal: $S_4$ ($\bar{4}$) 
			& 
			$
			\left(
			\begin{array}{ccc}
			0 & \alpha_2 & 0 \\
			\alpha_2 & 0 & 0 \\
			0     &    0     & 0
			\end{array}
			\right)
			$ 
			&
			Tetragonal: $D_{2d}$ ($\bar{4}2m$)
			\\
			\rule{0pt}{5ex}
			\multirow{3}{*}{
				$
				\left(
				\begin{array}{ccc}
				\alpha_0 & \alpha_2 & 0 \\
				-\alpha_2 & \alpha_0 & 0 \\
				0     &    0     & \alpha_1
				\end{array}
				\right)
				$              }
			& Tetragonal: $C_4$ ($4$) & 
			\multirow{3}{*}{
				$
				\left(
				\begin{array}{ccc}
				0 & \alpha_2 & 0 \\
				-\alpha_2 & 0 & 0 \\
				0     &    0     & 0
				\end{array}
				\right)
				$              }            & Tetragonal: $C_{4v}$ ($4mm$) \\ 
			& Trigonal: $C_3$ ($3$)   & & Trigonal:  $C_{3v}$ ($3m$)  \\
			& Hexagonal: $C_6$ ($6$)  & & Hexagonal: $C_{6v}$ ($6mm$) \\
			\rule{0pt}{5ex}
			\multirow{3}{*}{
				$
				\left(
				\begin{array}{ccc}
				\alpha_0 & 0 & 0 \\
				0 & \alpha_0 & 0 \\
				0     &    0     & \alpha_1
				\end{array}
				\right)
				$              } 
			& Tetragonal: $D_4$ ($422$)   & 
			\multirow{3}{*}{
				$
				\left(
				\begin{array}{ccc}
				\alpha_0 & 0 & 0 \\
				0 & \alpha_0 & 0 \\
				0     &    0     & \alpha_0
				\end{array}
				\right)
				$              }
			& \\ 
			& Trigonal: $D_3$ ($32$)      & & Cubic: $T$ ($23$), $O$ ($432$) \\
			& Hexagonal: $D_6$ ($622$)    & & \\
		\end{tabular}
	\end{ruledtabular}
\end{table*}


\subsection{Lack of duality symmetry and helicity}
In systems with low crystalline symmetry, the definition of
optical chirality meets with difficulties. For illustration, 
we consider a simple example of the non-gyrotropic system that belongs to the orthorhombic crystal
class. In this case, the electric permittivity and magnetic permeability
tensors in principal axes are given by diagonal matrices 
$\hat{\ee} = \diag(\ee_1, \ee_2, \ee_3)$ and $\hat{\mu} = \diag(\mu_1, \mu_2, \mu_3)$. 

For the diagonal $\hat{\ee}$ and $\hat{\mu}$, 
the symmetry analysis developed in Sec.~\ref{sec:gen} remains valid
along the principal axes. We consider one particular
direction taken as $\hat{z}$-axis, and fix $\bm{p} = p \hat{z}$.
To implement the general formalism in Eqs.~(\ref{eq:L1A})--(\ref{eq:H1}), we use the nonunitary 
transformation $\psi = \mathcal{N}_{2} \bar{\psi}$ ($\mathcal{N}_{2}$ is specified in 
Appendix~\ref{app:OR}), which leads to the diagonal form 
\begin{equation}
\bar{\mathcal{H}}^{(\mathcal{A})} =
\left(
\frac{-p}{\sqrt{\ee_1\mu_2}},   \frac{ p}{\sqrt{\ee_2\mu_1}}, 0
\frac{ p}{\sqrt{\ee_1\mu_2}},   \frac{-p}{\sqrt{\ee_2\mu_1}}, 0
\right),
\end{equation}
prescribed by Eq.~(\ref{eq:rot}).
In accordance with Eq.~(\ref{eq:[QH]}), the symmetry transformations in the transformed 
frame are given in Eqs.~(\ref{eq:QBrot})--(\ref{eq:QBrot2}).

However, neither duality nor helicity operators can be expressed in this basis. To illustrate this fact,
we apply the inverse transformation to the initial basis, $ \bar{\psi} = \mathcal{N}_{2}^{-1} \psi$, to the operators in Eqs.~(\ref{eq:QBrot})--(\ref{eq:QBrot2}). Straightforward calculation yields the following basis operators in the initial frame
\begin{align}
\label{eq:Qpr1}
\mathcal{Q}_{3}' = \rho \mathcal{N} \bar{\mathcal{Q}}_{a} \,
\mathcal{N}_{2}^{-1}, & \qquad
 \mathcal{Q}_{6}' = \rho \mathcal{N} \bar{\mathcal{Q}}_{b} \,
\mathcal{N}_{2}^{-1}, \\
\label{eq:Qpr2}
\mathcal{Q}_{7}' = \rho \mathcal{N} \bar{\mathcal{Q}}_{c} \,
\mathcal{N}_{2}^{-1}, & \qquad
 \mathcal{Q}_{8}' = \rho \mathcal{N} \bar{\mathcal{Q}}_{d} \,
\mathcal{N}_{2}^{-1},
\end{align}
where the explicit form of these matrices is given in Appendix~\ref{app:OR}.
Now if we take the limiting case of isotropic system, $\ee_i = \mu_i = 1$ ($i = 1,2$), we will find that Eqs.~(\ref{eq:Qpr1}, \ref{eq:Qpr2}) are mapped to the commutative subalgebra $\left\lbrace \mathcal{Q}_{3},\mathcal{Q}_{6}, \mathcal{Q}_{7},\mathcal{Q}_{8}\right\rbrace$ 
of $A_{8}$, see Eqs.~(\ref{eq:QA})--(\ref{eq:Q8}), which contains neither $ \mathcal{Q}_{2} $ (duality) nor $ \mathcal{Q}_{5} $ (helicity) operators.

From Eqs.~(\ref{eq:Qpr1}, \ref{eq:Qpr2}), we find that the only possible 
$ T $-even and $ P $-odd combinations are $ p_z \mathcal{Q}_6' $ and $ p_z \mathcal{Q}_7' $.
However, substitution of these expressions into Eq.~(\ref{eq:IOM3}) gives zero.
Therefore, we make a conclusion that it is not possible to construct optical chirality in the system, where the both duality and helicity are absent. This result is supported by the conclusions in Ref.~\cite{Fernandez2013a} that helicity-conserving scattering theorems do not exist for systems with one and two fold principal rotational axes.


\section{Summary}
\label{sec:sum}
We examined symmetry properties of the Maxwell's equations in various types of gyrotropic media 
making a particular accent on the conservation of optical chirality.  For this purpose, we extended the formalism of nongeometric symmetries in vacuum to medium with given constituent relations.  Within this approach, a conclusion about the conservation of optical chirality is reduced to the analysis of the invariance algebra of the nongeometric symmetries and establishing possible isomorphism between some elements of this algebra and operators of the helicity and duality symmetries in 
vacuum.  The advantage of this approach is that it suggests a straightforward way to derive various conservation laws related to the invariance algebra of the Maxwell's equations.

Using this method, we constructed the conservation law for optical chirality in isotropic chiral  media, as well as in trigonal, tetragonal, and hexagonal crystals along the symmetry direction. 
In particular, we demonstrated that in the gyrotropic crystals with  natural optical activity, which belong to the  point groups $C_n$ or $C_{nv}$,  only the optical helicity retains along the principal axis,  whereas in the case of achiral optically active crystals of the point symmetry $S_4$ or $D_{2d}$ only the duality transformation remains.
In all of the presented examples, except the achiral materials, we deal with a reduction of the original eight-dimensional invariance algebra in the vacuum to the four-dimensional
basis set. Additionally, we give an example of the medium where none of these symmetries is conserved.


\begin{acknowledgments}
	We are thankful to K.~Y.~Bliokh for useful comments.
    The work was supported by the Government of the 
    Russian Federation Program 02.A03.21.0006 
    and by the Ministry of Education and Science of the Russian Federation, 
    projects Nos. 1437 and 2725.
	The authors also acknowledge support by 
    JSPS KAKENHI Grants No. 25287087 and No. 25220803.
    I.~P. acknowledges financial support by Center for Chiral
    Science, Hiroshima University and by Ministry of Education and
    Science of the Russian Federation, Grant No.~MK-6230.2016.2.
\end{acknowledgments}


\appendix

\onecolumngrid

\section{Derivation of the invariance algebra in medium}
\label{app:IA}

We highlight the derivation of the invariance algebra for the Maxwell's equations in (\ref{eq:Max1}, \ref{eq:Max2}).  Any symmetry transformation $\mathcal{Q}_{A}(\bm{p})$ that transforms a solution $\psi(t, \bm{p}) = (\bm{D}(t, \bm{p}), \bm{B}(t, \bm{p}))$ of Eqs.~(\ref{eq:Max1}, \ref{eq:Max2}) into another solution $\psi' = \mathcal{Q}_{A} \psi$ should satisfy the following invariance conditions
\begin{eqnarray}
\label{eq:invar3}
[\mathcal{L}^{(\mathcal{A})}, \mathcal{Q}_A ] &=& g_{11}
\mathcal{L}^{(\mathcal{A})} + g_{12} \mathcal{L},     \\
\label{eq:invar4}
\left [\mathcal{L}, \mathcal{Q}_A \right] &=& g_{21}
\mathcal{L}^{(\mathcal{A})} + g_{22} \mathcal{L},
\end{eqnarray}
where $\mathcal{L}$ and $ \mathcal{L}^{(\mathcal{A})} $ are determined in Eqs.~(\ref{eq:L3}) and (\ref{eq:L1A}), and $g_{ij}$ denote some arbitrary operators acting on $\psi$.

For the diagonal operators in
Eqs.~(\ref{eq:rot}, \ref{eq:rot11}), we can find the invariance algebra. 
Given the fact that ${\cal Q}_A(\bm p)$ depends only on $\bm{p}$, 
the invariance conditions in the transformed frame take the following reduced form
\begin{eqnarray}
\label{eq:inv2}
[\bar{{\cal H}}^{(\mathcal{A})}, \bar{\cal Q}_A]  &=&  \bar{g}_{12}
\bar{\mathcal{L}},\\
\label{eq:inv22}
\left[\bar{{\cal L}}, \bar{\cal Q}_A \right]  &=&  \bar{g}_{22} \bar{{\cal L}},
\end{eqnarray}
where $\bar{{\cal Q}}_A  =  \mathcal{N}^{-1} {\cal Q}_A \mathcal{N}$, 
and  $\bar{g}_{12}$ and $\bar{g}_{22}$ denote some redefined operators.

The most general form of $ \bar{\cal Q}_A$ imposed by
Eq.~(\ref{eq:inv22}) is rendered as 
\begin{equation}
\label{eq:Qa}
\bar{\cal Q}_A = \left(
\begin{array}{cccccc}
q_{11} & q_{12} &   0   & q_{14} & q_{15} &   0     \\
q_{21} & q_{22} &   0   & q_{24} & q_{25} &   0     \\
0  &    0  &   0   &    0  &    0  &   0     \\
q_{41} & q_{42} &   0   & q_{44} & q_{45} &   0     \\
q_{51} & q_{52} &   0   & q_{54} & q_{55} &   0     \\
0  &    0  &   0   &    0  &    0  &   0
\end{array}
\right) + \mathcal{F} \bar{\mathcal{L}},
\end{equation}
with some operator $\mathcal{F}$. 
The last term in  this equation can be safely dropped since it 
does not contribute into finding of the invariance algebra \cite{Fushchich1987}.
Then, the first invariance condition given by Eq.~(\ref{eq:inv2})
is satisfied for commutative $\bar{{\cal H}}^{(\mathcal{A})}$
and $\bar{\cal Q}_A$.

To identify the invariance algebra, we explicitly calculate the commutator in Eq.~(\ref{eq:inv2})
	\begin{equation}
	\label{eq:[QH]}
	\left[\bar{{\cal H}}^{(\mathcal{A})}, \bar{\cal Q}_A\right]
	= \left(
	\begin{array}{cccccc}
	0  & q_{12}(\omega_2 - \omega_1) &   0   & q_{14}(\omega_4 - \omega_1) & q_{15}(\omega_5- \omega_1) &   0     \\
	q_{21}(\omega_1 - \omega_2) &    0  &   0   & q_{24}(\omega_4 - \omega_2) & q_{25}(\omega_5 - \omega_2) &   0     \\
	0  &    0  &   0   &    0  &    0  &   0     \\
	q_{41}(\omega_1 - \omega_4) & q_{42}(\omega_2 - \omega_4) &   0   &    0  & q_{45}(\omega_5 -\omega_4) &   0     \\
	q_{51}(\omega_1 - \omega_5) & q_{52}(\omega_2 - \omega_5) &   0   & q_{54}(\omega_4 - \omega_5) &    0  &   0     \\
	0  &    0  &   0   &    0  &    0  &   0
	\end{array}
	\right),
	\end{equation}
where Eq.~(\ref{eq:rot}) is used. The number of basis elements in the invariance
algebra depends on the symmetry relations between $\omega_i$ that occur in Eq.~(\ref{eq:rot}).
When there are no any degeneracies between $\omega_i$  (the lowest symmetry case), 
all the matrices $\bar{\cal Q}_A$, which commute with $\bar{{\cal H}}^{(\mathcal{A})}$, have diagonal form with four free parameters
\begin{equation}
\label{eq:Q4}
\bar{\cal Q}_A = \diag(q_{11}, q_{22}, 0, q_{44}, q_{55}, 0),
\end{equation}
which corresponds to the set of basis operators in Eq.~(\ref{eq:Qb}--\ref{eq:QBrot2}).

\section{Transformations in point groups $C_{n}$, $C_{nv}$, and $D_{n}$, $n \ge 3$}
\label{app:C4}

In the point groups $C_{n}$, $C_{nv}$, and $D_{n}$, the inverse tensors  $\hat{\ee}^{-1}$
and $\hat{\mu}^{-1}$ in Eq.~(\ref{eq:H1}) are given by
\begin{align}
\hat{\ee}^{-1} = \frac{1}{\ee_\perp d} \left(
\begin{array}{ccc}
1 + i \alpha_2 p & i \alpha_0 p & 0 \\
-i \alpha_0 p & 1 + i\alpha_2 p & 0 \\
0 & 0 & 0 
\end{array}
\right) + \frac{1}{\ee_\parallel}
\left(
\begin{array}{ccc}
0 & 0 & 0 \\
0 & 0 & 0 \\
0 & 0 & 1
\end{array}
\right),
\\
\hat{\mu}^{-1} = \frac{1}{\mu_\perp d^*} \left(
\begin{array}{ccc}
1 - i \alpha_2 p & i \alpha_0 p & 0 \\
-i \alpha_0 p & 1 - i\alpha_2 p & 0 \\
0 & 0 & 0 
\end{array}
\right) + \frac{1}{\mu_\parallel}
\left(
\begin{array}{ccc}
0 & 0 & 0 \\
0 & 0 & 0 \\
0 & 0 & 1
\end{array}
\right),
\end{align}
where $d = 1 - |a|^2 + 2 i \alpha_2 p $ and $a = (\alpha_0 + i \alpha_2) p$.
In what follows, we will use the units where $\ee_\perp = \mu_\perp =1$.

The matrix $\mathcal{H}^{(\mathcal{A})}$ in Eq.~(\ref{eq:H1}) is diagonalized 
in two steps. First, we make a transformation to the helicity basis
\begin{equation}
\mathcal{H}^{(\mathcal{A})}_{\Lambda} =  \mathcal{U}^{\dag}_{\Lambda} 
\mathcal{H}^{(\mathcal{A})} \mathcal{U}_{\Lambda},
\end{equation}
where $\mathcal{U}_{\Lambda} = \sigma_{0} \otimes \hat{U}_{\Lambda}$, see Eq.~(\ref{eq:UL}),
which gives
\begin{equation}
\mathcal{H}^{(\mathcal{A})}_{\Lambda} = 
\begin{pmatrix}
0 & i \hat{M}^{*} \\
- i\hat{M} & 0 
\end{pmatrix},
\end{equation}
where $\hat{M} = \diag\left(-p/(1 + a^{*}), p/(1 + a), 0 \right)$.
Second, we apply a unitary transformation 
\begin{equation}
\bar{\mathcal{H}}^{(\mathcal{A})} =  \mathcal{U}_{M}^{\dag}
\mathcal{H}^{(\mathcal{A})}_{\Lambda} \mathcal{U}_{M},
\end{equation}
with
\begin{equation}
\mathcal{U}_M = \frac{1}{\sqrt{2}}\left(
\begin{array}{cccccc}
- \dfrac{i p}{p_{+}(1 - a)} & 0 & 0  & \dfrac{i p}{p_{+}(1 - a)} & 0 & 0 \\
0 & -\dfrac{i p}{p_{-}(1 + a^*)} & 0 & 0 & \dfrac{i p}{p_{-}(1 + a^*)} & 0 \\
0 & 0 & \sqrt{2} & 0 & 0 & 0 \\
1 & 0 & 0  & 1 & 0 & 0 \\
0 & 1  & 0 & 0 & 1 & 0 \\
0 & 0 & 0 & 0 & 0 & \sqrt{2}  
\end{array}
\right),
\end{equation}
which gives the diagonal form Eq.~(\ref{eq:Ha1}). 
Altogether, the whole transformation in Eq.~(\ref{eq:UC}) is rendered by
\begin{equation}
\mathcal{U}_{C} = \mathcal{U}_{\Lambda} \mathcal{U}_{M}.
\end{equation}

Let us note that the conservation law for the optical chirality in Eq.~(\ref{eq:100})
is obtained directly, if we notice that the  inverse transformation of  
$\bar{\mathcal{Q}}_{b}$ in Eq.~(\ref{eq:Qb}) gives the helicity operator
\begin{equation}
(\bm{\mathcal{S}} \cdot \tilde{\bm{p}}) = \mathcal{U}_{C}\, \bar{\mathcal{Q}}_{b} \,
\mathcal{U}_{C}^{\dag},
\end{equation}
then optical chirality can be defined similar to Eq.~(\ref{eq:Zvac})
\begin{equation}\label{eq:A9}
C_{\chi}^{(c)} = \frac{1}{2} \int d^3 p \,
\psi^{\dag}(t, \bm{p})\, (\bm{\mathcal{S}} \cdot \bm{p})\, \psi(t, \bm{p}),
\end{equation}
which corresponds to Eq.~(\ref{eq:100}).

\section{Transformations in point groups $S_{4}$ and $D_{2d}$}
\label{app:S4}

In achiral point groups $S_{4}$ and $D_{2d}$, we have the following  
$\hat{\ee}^{-1}$ and $\hat{\mu}^{-1}$ in Eq.~(\ref{eq:H1})
\begin{align}
\hat{\ee}^{-1} = \frac{1}{\ee_\perp(1 + |a|^2)} \left(
\begin{array}{ccc}
1 - i \alpha_2 p & i \alpha_0 p & 0 \\
i \alpha_0 p & 1 + i \alpha_2 p & 0 \\
0 & 0 & 0
\end{array}
\right) +
\frac{1}{\ee_\parallel} \left(
\begin{array}{ccc}
0 & 0 & 0 \\
0 & 0 & 0 \\
0 & 0 & 1
\end{array}
\right),
\\
\hat{\mu}^{-1} = \frac{1}{\mu_\perp(1 + |a|^2)} \left(
\begin{array}{ccc}
1 - i \alpha_2 p & i \alpha_0 p & 0 \\
i \alpha_0 p & 1 + i \alpha_2 p & 0 \\
0 & 0 & 0
\end{array}
\right) +
\frac{1}{\mu_\parallel} \left(
\begin{array}{ccc}
0 & 0 & 0 \\
0 & 0 & 0 \\
0 & 0 & 1
\end{array}
\right).
\end{align}

In order to diagonalize $\mathcal{H}^{(\mathcal{A})}$ in Eq.~(\ref{eq:H1}),
we apply a sequence of unitary transformations
\begin{equation}
\mathcal{U}_{S} = \mathcal{U}_{\Lambda} \mathcal{U}_{N} \mathcal{U}_{2},
\end{equation}
where $\mathcal{U}_{\Lambda} = \sigma_{0} \otimes \hat{U}_{\Lambda}$, $\mathcal{U}_{N} = 
\sigma_{0} \otimes \hat{U}_{N}$, and $\mathcal{U}_{2} = \hat{U}_{2} \otimes{I}$  with 
\begin{equation}
\hat{U}_{\Lambda} = \left(
\begin{array}{ccc}
-\dfrac{1}{\sqrt{2}} & \dfrac{1}{\sqrt{2}} & 0 \\
\dfrac{i}{\sqrt{2}} & \dfrac{i}{\sqrt{2}} & 0 \\
0 & 0 & 1
\end{array}
\right),
\qquad
\hat{U}_N = 
\left(
\begin{array}{ccc}
\dfrac{|a|}{\sqrt{2} a^*} \dfrac{1-\sqrt{1 + |a|^2}}{\sqrt{1 + |a|^2 -\sqrt{1 + |a|^2} }} & 
\dfrac{|a|}{\sqrt{2} a^*} \dfrac{1+\sqrt{1 + |a|^2}}{\sqrt{1 + |a|^2 +\sqrt{1 + |a|^2} }} & 0 \\
\dfrac{|a|}{\sqrt{2}} \dfrac{1}{\sqrt{1 + |a|^2 -\sqrt{1 + |a|^2} }} &
\dfrac{|a|}{\sqrt{2}} \dfrac{1}{\sqrt{1 + |a|^2 +\sqrt{1 + |a|^2} }} & 0 \\
0 & 0 & 1
\end{array}
\right).
\end{equation}
Note that $\mathcal{U}_{N}$ commutes with 
$\bar{\mathcal{L}} 
= \mathcal{U}^{\dag}_{\Lambda}\,\mathcal{L}\, \mathcal{U}_{\Lambda}$.

\section{Orthorhombic crystal}
\label{app:OR}

The diagonalization in Eq.~(\ref{eq:rot}) is carried out by applying 
\begin{equation}
\mathcal{N} = 
\begin{pmatrix}
-\sqrt{\frac{\ee_1}{\ee_1 + \mu_2}} & 0 & 0 &\sqrt{\frac{\ee_1}{\ee_1 + \mu_2}} & 0 & 0 \\
0 & -\sqrt{\frac{\ee_2}{\ee_2 + \mu_1}} &0&0 &\sqrt{\frac{\ee_2}{\ee_2 + \mu_1}} &0 \\
0 & 0 & 1 & 0 & 0 & 0 \\ 
0 & \sqrt{\frac{\mu_1}{\mu_1 + \ee_2}} & 0 & 0 & \sqrt{\frac{\mu_1}{\mu_1 + \ee_2}} & 0\\
\sqrt{\frac{\mu_2}{\mu_2 + \ee_1}} & 0 & 0 & \sqrt{\frac{\mu_2}{\mu_2 + \ee_1}} & 0&0 \\
0 & 0 & 0 & 0 & 0 & 1 
\end{pmatrix}
\end{equation}
that commutes with $\mathcal{L}$ in Eq.~(\ref{eq:L3}).

The explicit form of  $\mathcal{Q}_{A}'$ in Eqs.~(\ref{eq:Qpr1}, \ref{eq:Qpr2})
is given by the following expressions
\begin{align}
\mathcal{Q}_{3}' = \left(
\begin{array}{cccccc}
0 & 0 & 0 & 0 &  -\frac{\ee_1 + \mu_2}{2\sqrt{\ee_1\mu_2}} & 0 \\
0 & 0&0 & -\frac{\ee_2 + \mu_1}{2\sqrt{\ee_2\mu_1}}&0 & 0\\
0 & 0 & 0 & 0 & 0 & 0 \\ 
0 & -\frac{\ee_2 + \mu_1}{2\sqrt{\ee_2\mu_1}} & 0 & 0 & 0 & 0\\
-\frac{\ee_1 + \mu_2}{2\sqrt{\ee_1\mu_2}} & 0 & 0 & 0 & 0&0 \\
0 & 0 & 0 & 0 & 0 & 0
\end{array}
\right),
& \quad
\mathcal{Q}_{8}'  = \left(
\begin{array}{cccccc}
0 & 0 & 0 & 0 &  \frac{\ee_1 + \mu_2}{2\sqrt{\ee_1\mu_2}} & 0 \\
0 & 0&0 & -\frac{\ee_2 + \mu_1}{2\sqrt{\ee_2\mu_1}}&0 & 0\\
0 & 0 & 0 & 0 & 0 & 0 \\ 
0 & -\frac{\ee_2 + \mu_1}{2\sqrt{\ee_2\mu_1}} & 0 & 0 & 0 & 0\\
\frac{\ee_1 + \mu_2}{2\sqrt{\ee_1\mu_2}} & 0 & 0 & 0 & 0&0 \\
0 & 0 & 0 & 0 & 0 & 0
\end{array}
\right), \\
\mathcal{Q}_{6}'  = \diag \left(
-\frac{\ee_1 + \mu_2}{2\ee_1} ,
\frac{\ee_2 + \mu_1}{2\ee_2}, 
0,
\frac{\ee_2 + \mu_1}{2\mu_1},
-\frac{\ee_1 + \mu_2}{2\mu_2}, 
0
\right),
& \quad
\mathcal{Q}_{7}' = \left(
\frac{\ee_1 + \mu_2}{2\ee_1}, 
\frac{\ee_2 + \mu_1}{2\ee_2},
0,
\frac{\ee_2 + \mu_1}{2\mu_1},
\frac{\ee_1 + \mu_2}{2\mu_2},
0
\right).
\end{align}

\twocolumngrid


\bibliography{maxwell}

\begin{thebibliography}{62}%
\makeatletter
\providecommand \@ifxundefined [1]{%
 \@ifx{#1\undefined}
}%
\providecommand \@ifnum [1]{%
 \ifnum #1\expandafter \@firstoftwo
 \else \expandafter \@secondoftwo
 \fi
}%
\providecommand \@ifx [1]{%
 \ifx #1\expandafter \@firstoftwo
 \else \expandafter \@secondoftwo
 \fi
}%
\providecommand \natexlab [1]{#1}%
\providecommand \enquote  [1]{``#1''}%
\providecommand \bibnamefont  [1]{#1}%
\providecommand \bibfnamefont [1]{#1}%
\providecommand \citenamefont [1]{#1}%
\providecommand \href@noop [0]{\@secondoftwo}%
\providecommand \href [0]{\begingroup \@sanitize@url \@href}%
\providecommand \@href[1]{\@@startlink{#1}\@@href}%
\providecommand \@@href[1]{\endgroup#1\@@endlink}%
\providecommand \@sanitize@url [0]{\catcode `\\12\catcode `\$12\catcode
  `\&12\catcode `\#12\catcode `\^12\catcode `\_12\catcode `\%12\relax}%
\providecommand \@@startlink[1]{}%
\providecommand \@@endlink[0]{}%
\providecommand \url  [0]{\begingroup\@sanitize@url \@url }%
\providecommand \@url [1]{\endgroup\@href {#1}{\urlprefix }}%
\providecommand \urlprefix  [0]{URL }%
\providecommand \Eprint [0]{\href }%
\providecommand \doibase [0]{http://dx.doi.org/}%
\providecommand \selectlanguage [0]{\@gobble}%
\providecommand \bibinfo  [0]{\@secondoftwo}%
\providecommand \bibfield  [0]{\@secondoftwo}%
\providecommand \translation [1]{[#1]}%
\providecommand \BibitemOpen [0]{}%
\providecommand \bibitemStop [0]{}%
\providecommand \bibitemNoStop [0]{.\EOS\space}%
\providecommand \EOS [0]{\spacefactor3000\relax}%
\providecommand \BibitemShut  [1]{\csname bibitem#1\endcsname}%
\let\auto@bib@innerbib\@empty
\bibitem [{\citenamefont {Kelvin}(1904)}]{Kelvin1904}%
  \BibitemOpen
  \bibfield  {author} {\bibinfo {author} {\bibfnamefont {L.}~\bibnamefont
  {Kelvin}},\ }\href@noop {} {\emph {\bibinfo {title} {Baltimore Lectures on
  Molecular Dynamics and the Wave Theory of Light}}}\ (\bibinfo  {publisher}
  {CJ Clay \& Sons, London},\ \bibinfo {year} {1904})\BibitemShut {NoStop}%
\bibitem [{\citenamefont {Dreiling}\ and\ \citenamefont
  {Gay}(2014)}]{Dreiling2014}%
  \BibitemOpen
  \bibfield  {author} {\bibinfo {author} {\bibfnamefont {J.~M.}\ \bibnamefont
  {Dreiling}}\ and\ \bibinfo {author} {\bibfnamefont {T.~J.}\ \bibnamefont
  {Gay}},\ }\href {\doibase 10.1103/PhysRevLett.113.118103} {\bibfield
  {journal} {\bibinfo  {journal} {Phys. Rev. Lett.}\ }\textbf {\bibinfo
  {volume} {113}},\ \bibinfo {pages} {118103} (\bibinfo {year}
  {2014})}\BibitemShut {NoStop}%
\bibitem [{\citenamefont {Lipkin}(1964)}]{Lipkin1964}%
  \BibitemOpen
  \bibfield  {author} {\bibinfo {author} {\bibfnamefont {D.~M.}\ \bibnamefont
  {Lipkin}},\ }\href {\doibase http://dx.doi.org/10.1063/1.1704165} {\bibfield
  {journal} {\bibinfo  {journal} {J. Math. Phys.}\ }\textbf {\bibinfo {volume}
  {5}},\ \bibinfo {pages} {696} (\bibinfo {year} {1964})}\BibitemShut {NoStop}%
\bibitem [{\citenamefont {Morgan}(1964)}]{Morgan1964}%
  \BibitemOpen
  \bibfield  {author} {\bibinfo {author} {\bibfnamefont {T.~A.}\ \bibnamefont
  {Morgan}},\ }\href {\doibase http://dx.doi.org/10.1063/1.1931204} {\bibfield
  {journal} {\bibinfo  {journal} {J. Math. Phys.}\ }\textbf {\bibinfo {volume}
  {5}},\ \bibinfo {pages} {1659} (\bibinfo {year} {1964})}\BibitemShut
  {NoStop}%
\bibitem [{\citenamefont {Kibble}(1965)}]{Kibble1964}%
  \BibitemOpen
  \bibfield  {author} {\bibinfo {author} {\bibfnamefont {T.~W.~B.}\
  \bibnamefont {Kibble}},\ }\href {\doibase
  http://dx.doi.org/10.1063/1.1704363} {\bibfield  {journal} {\bibinfo
  {journal} {J. Math. Phys.}\ }\textbf {\bibinfo {volume} {6}},\ \bibinfo
  {pages} {1022} (\bibinfo {year} {1965})}\BibitemShut {NoStop}%
\bibitem [{\citenamefont {Barron}(1986)}]{Barron1986}%
  \BibitemOpen
  \bibfield  {author} {\bibinfo {author} {\bibfnamefont {L.}~\bibnamefont
  {Barron}},\ }\href {\doibase http://dx.doi.org/10.1016/0009-2614(86)80035-5}
  {\bibfield  {journal} {\bibinfo  {journal} {Chem. Phys. Lett.}\ }\textbf
  {\bibinfo {volume} {123}},\ \bibinfo {pages} {423 } (\bibinfo {year}
  {1986})}\BibitemShut {NoStop}%
\bibitem [{\citenamefont {Barron}(2004)}]{Barron2004}%
  \BibitemOpen
  \bibfield  {author} {\bibinfo {author} {\bibfnamefont {L.~D.}\ \bibnamefont
  {Barron}},\ }\href@noop {} {\emph {\bibinfo {title} {Molecular light
  scattering and optical activity}}}\ (\bibinfo  {publisher} {Cambridge
  University Press},\ \bibinfo {year} {2004})\BibitemShut {NoStop}%
\bibitem [{\citenamefont {Fushchich}\ and\ \citenamefont
  {Nikitin}(1987)}]{Fushchich1987}%
  \BibitemOpen
  \bibfield  {author} {\bibinfo {author} {\bibfnamefont {W.~I.}\ \bibnamefont
  {Fushchich}}\ and\ \bibinfo {author} {\bibfnamefont {A.~G.}\ \bibnamefont
  {Nikitin}},\ }\href@noop {} {\emph {\bibinfo {title} {Symmetries of
  Maxwell’s Equations}}},\ Mathematics and its Applications\ (\bibinfo
  {publisher} {Springer Netherlands},\ \bibinfo {year} {1987})\BibitemShut
  {NoStop}%
\bibitem [{\citenamefont {Krivskii}\ and\ \citenamefont
  {Simulik}(1989{\natexlab{a}})}]{Krivskii1989a}%
  \BibitemOpen
  \bibfield  {author} {\bibinfo {author} {\bibfnamefont {I.~Y.}\ \bibnamefont
  {Krivskii}}\ and\ \bibinfo {author} {\bibfnamefont {V.~M.}\ \bibnamefont
  {Simulik}},\ }\href {\doibase 10.1007/BF01016113} {\bibfield  {journal}
  {\bibinfo  {journal} {Theor. Math. Phys.}\ }\textbf {\bibinfo {volume}
  {80}},\ \bibinfo {pages} {864} (\bibinfo {year}
  {1989}{\natexlab{a}})}\BibitemShut {NoStop}%
\bibitem [{\citenamefont {Krivskii}\ and\ \citenamefont
  {Simulik}(1989{\natexlab{b}})}]{Krivskii1989b}%
  \BibitemOpen
  \bibfield  {author} {\bibinfo {author} {\bibfnamefont {I.~Y.}\ \bibnamefont
  {Krivskii}}\ and\ \bibinfo {author} {\bibfnamefont {V.~M.}\ \bibnamefont
  {Simulik}},\ }\href {\doibase 10.1007/BF01016183} {\bibfield  {journal}
  {\bibinfo  {journal} {Theor. Math. Phys.}\ }\textbf {\bibinfo {volume}
  {80}},\ \bibinfo {pages} {912} (\bibinfo {year}
  {1989}{\natexlab{b}})}\BibitemShut {NoStop}%
\bibitem [{\citenamefont {Ibragimov}(2008)}]{Ibragimov2008}%
  \BibitemOpen
  \bibfield  {author} {\bibinfo {author} {\bibfnamefont {N.~H.}\ \bibnamefont
  {Ibragimov}},\ }\href {\doibase 10.1007/s10440-008-9270-y} {\bibfield
  {journal} {\bibinfo  {journal} {Acta Appl. Math.}\ }\textbf {\bibinfo
  {volume} {105}},\ \bibinfo {pages} {157} (\bibinfo {year}
  {2008})}\BibitemShut {NoStop}%
\bibitem [{\citenamefont {Cameron}\ and\ \citenamefont
  {Barnett}(2012)}]{Cameron2012}%
  \BibitemOpen
  \bibfield  {author} {\bibinfo {author} {\bibfnamefont {R.~P.}\ \bibnamefont
  {Cameron}}\ and\ \bibinfo {author} {\bibfnamefont {S.~M.}\ \bibnamefont
  {Barnett}},\ }\href {http://stacks.iop.org/1367-2630/14/i=12/a=123019}
  {\bibfield  {journal} {\bibinfo  {journal} {New J. Phys.}\ }\textbf {\bibinfo
  {volume} {14}},\ \bibinfo {pages} {123019} (\bibinfo {year}
  {2012})}\BibitemShut {NoStop}%
\bibitem [{\citenamefont {Bliokh}\ \emph {et~al.}(2013)\citenamefont {Bliokh},
  \citenamefont {Bekshaev},\ and\ \citenamefont {Nori}}]{Bliokh2013}%
  \BibitemOpen
  \bibfield  {author} {\bibinfo {author} {\bibfnamefont {K.~Y.}\ \bibnamefont
  {Bliokh}}, \bibinfo {author} {\bibfnamefont {A.~Y.}\ \bibnamefont
  {Bekshaev}}, \ and\ \bibinfo {author} {\bibfnamefont {F.}~\bibnamefont
  {Nori}},\ }\href {http://stacks.iop.org/1367-2630/15/i=3/a=033026} {\bibfield
   {journal} {\bibinfo  {journal} {New J. Phys.}\ }\textbf {\bibinfo {volume}
  {15}},\ \bibinfo {pages} {033026} (\bibinfo {year} {2013})}\BibitemShut
  {NoStop}%
\bibitem [{\citenamefont {Cameron}(2014)}]{Cameron2014}%
  \BibitemOpen
  \bibfield  {author} {\bibinfo {author} {\bibfnamefont {R.~P.}\ \bibnamefont
  {Cameron}},\ }\emph {\bibinfo {title} {On the angular momentum of light}},\
  \href@noop {} {Ph.D. thesis},\ \bibinfo  {school} {School of Physics and
  Astronomy, College of Science of Engineering, University of Glasgow}
  (\bibinfo {year} {2014})\BibitemShut {NoStop}%
\bibitem [{\citenamefont {Barnett}(2014)}]{Barnett2014}%
  \BibitemOpen
  \bibfield  {author} {\bibinfo {author} {\bibfnamefont {S.~M.}\ \bibnamefont
  {Barnett}},\ }\href {http://stacks.iop.org/1367-2630/16/i=9/a=093008}
  {\bibfield  {journal} {\bibinfo  {journal} {New J. Phys.}\ }\textbf {\bibinfo
  {volume} {16}},\ \bibinfo {pages} {093008} (\bibinfo {year}
  {2014})}\BibitemShut {NoStop}%
\bibitem [{\citenamefont {Philbin}(2013)}]{Philbin2013}%
  \BibitemOpen
  \bibfield  {author} {\bibinfo {author} {\bibfnamefont {T.~G.}\ \bibnamefont
  {Philbin}},\ }\href {\doibase 10.1103/PhysRevA.87.043843} {\bibfield
  {journal} {\bibinfo  {journal} {Phys. Rev. A}\ }\textbf {\bibinfo {volume}
  {87}},\ \bibinfo {pages} {043843} (\bibinfo {year} {2013})}\BibitemShut
  {NoStop}%
\bibitem [{\citenamefont {Calkin}(1965)}]{Calkin1965}%
  \BibitemOpen
  \bibfield  {author} {\bibinfo {author} {\bibfnamefont {M.~G.}\ \bibnamefont
  {Calkin}},\ }\href {\doibase http://dx.doi.org/10.1119/1.1971089} {\bibfield
  {journal} {\bibinfo  {journal} {American Journal of Physics}\ }\textbf
  {\bibinfo {volume} {33}},\ \bibinfo {pages} {958} (\bibinfo {year}
  {1965})}\BibitemShut {NoStop}%
\bibitem [{\citenamefont {Zwanziger}(1968)}]{Zwanziger1968}%
  \BibitemOpen
  \bibfield  {author} {\bibinfo {author} {\bibfnamefont {D.}~\bibnamefont
  {Zwanziger}},\ }\href {\doibase 10.1103/PhysRev.176.1489} {\bibfield
  {journal} {\bibinfo  {journal} {Phys. Rev.}\ }\textbf {\bibinfo {volume}
  {176}},\ \bibinfo {pages} {1489} (\bibinfo {year} {1968})}\BibitemShut
  {NoStop}%
\bibitem [{\citenamefont {Drummond}(1999)}]{Drummond1999}%
  \BibitemOpen
  \bibfield  {author} {\bibinfo {author} {\bibfnamefont {P.~D.}\ \bibnamefont
  {Drummond}},\ }\href@noop {} {\bibfield  {journal} {\bibinfo  {journal}
  {Physical Review A}\ }\textbf {\bibinfo {volume} {60}},\ \bibinfo {pages}
  {R3331} (\bibinfo {year} {1999})}\BibitemShut {NoStop}%
\bibitem [{\citenamefont {Drummond}(2006)}]{Drummond2006}%
  \BibitemOpen
  \bibfield  {author} {\bibinfo {author} {\bibfnamefont {P.~D.}\ \bibnamefont
  {Drummond}},\ }\href@noop {} {\bibfield  {journal} {\bibinfo  {journal}
  {Journal of Physics B: Atomic, Molecular and Optical Physics}\ }\textbf
  {\bibinfo {volume} {39}},\ \bibinfo {pages} {S573} (\bibinfo {year}
  {2006})}\BibitemShut {NoStop}%
\bibitem [{\citenamefont {Tang}\ and\ \citenamefont {Cohen}(2010)}]{Tang2010}%
  \BibitemOpen
  \bibfield  {author} {\bibinfo {author} {\bibfnamefont {Y.}~\bibnamefont
  {Tang}}\ and\ \bibinfo {author} {\bibfnamefont {A.~E.}\ \bibnamefont
  {Cohen}},\ }\href {\doibase 10.1103/PhysRevLett.104.163901} {\bibfield
  {journal} {\bibinfo  {journal} {Phys. Rev. Lett.}\ }\textbf {\bibinfo
  {volume} {104}},\ \bibinfo {pages} {163901} (\bibinfo {year}
  {2010})}\BibitemShut {NoStop}%
\bibitem [{\citenamefont {Hendry}\ \emph {et~al.}(2010)\citenamefont {Hendry},
  \citenamefont {Carpy}, \citenamefont {Johnston}, \citenamefont {Popland},
  \citenamefont {Mikhaylovskiy}, \citenamefont {Lapthorn}, \citenamefont
  {Kelly}, \citenamefont {Barron}, \citenamefont {Gadegaard},\ and\
  \citenamefont {Kadodwala}}]{Hendry2010}%
  \BibitemOpen
  \bibfield  {author} {\bibinfo {author} {\bibfnamefont {E.}~\bibnamefont
  {Hendry}}, \bibinfo {author} {\bibfnamefont {T.}~\bibnamefont {Carpy}},
  \bibinfo {author} {\bibfnamefont {J.}~\bibnamefont {Johnston}}, \bibinfo
  {author} {\bibfnamefont {M.}~\bibnamefont {Popland}}, \bibinfo {author}
  {\bibfnamefont {R.}~\bibnamefont {Mikhaylovskiy}}, \bibinfo {author}
  {\bibfnamefont {A.}~\bibnamefont {Lapthorn}}, \bibinfo {author}
  {\bibfnamefont {S.}~\bibnamefont {Kelly}}, \bibinfo {author} {\bibfnamefont
  {L.}~\bibnamefont {Barron}}, \bibinfo {author} {\bibfnamefont
  {N.}~\bibnamefont {Gadegaard}}, \ and\ \bibinfo {author} {\bibfnamefont
  {M.}~\bibnamefont {Kadodwala}},\ }\href@noop {} {\bibfield  {journal}
  {\bibinfo  {journal} {Nature Nanotech.}\ }\textbf {\bibinfo {volume} {5}},\
  \bibinfo {pages} {783} (\bibinfo {year} {2010})}\BibitemShut {NoStop}%
\bibitem [{\citenamefont {Tang}\ and\ \citenamefont {Cohen}(2011)}]{Tang2011}%
  \BibitemOpen
  \bibfield  {author} {\bibinfo {author} {\bibfnamefont {Y.}~\bibnamefont
  {Tang}}\ and\ \bibinfo {author} {\bibfnamefont {A.~E.}\ \bibnamefont
  {Cohen}},\ }\href {\doibase 10.1126/science.1202817} {\bibfield  {journal}
  {\bibinfo  {journal} {Science}\ }\textbf {\bibinfo {volume} {332}},\ \bibinfo
  {pages} {333} (\bibinfo {year} {2011})}\BibitemShut {NoStop}%
\bibitem [{\citenamefont {Hendry}\ \emph {et~al.}(2012)\citenamefont {Hendry},
  \citenamefont {Mikhaylovskiy}, \citenamefont {Barron}, \citenamefont
  {Kadodwala},\ and\ \citenamefont {Davis}}]{Hendry2012}%
  \BibitemOpen
  \bibfield  {author} {\bibinfo {author} {\bibfnamefont {E.}~\bibnamefont
  {Hendry}}, \bibinfo {author} {\bibfnamefont {R.~V.}\ \bibnamefont
  {Mikhaylovskiy}}, \bibinfo {author} {\bibfnamefont {L.~D.}\ \bibnamefont
  {Barron}}, \bibinfo {author} {\bibfnamefont {M.}~\bibnamefont {Kadodwala}}, \
  and\ \bibinfo {author} {\bibfnamefont {T.~J.}\ \bibnamefont {Davis}},\ }\href
  {\doibase 10.1021/nl3012787} {\bibfield  {journal} {\bibinfo  {journal} {Nano
  Lett.}\ }\textbf {\bibinfo {volume} {12}},\ \bibinfo {pages} {3640} (\bibinfo
  {year} {2012})}\BibitemShut {NoStop}%
\bibitem [{\citenamefont {Kamenetskii}\ \emph {et~al.}(2013)\citenamefont
  {Kamenetskii}, \citenamefont {Joffe},\ and\ \citenamefont
  {Shavit}}]{Kamenetskii2013}%
  \BibitemOpen
  \bibfield  {author} {\bibinfo {author} {\bibfnamefont {E.~O.}\ \bibnamefont
  {Kamenetskii}}, \bibinfo {author} {\bibfnamefont {R.}~\bibnamefont {Joffe}},
  \ and\ \bibinfo {author} {\bibfnamefont {R.}~\bibnamefont {Shavit}},\ }\href
  {\doibase 10.1103/PhysRevE.87.023201} {\bibfield  {journal} {\bibinfo
  {journal} {Phys. Rev. E}\ }\textbf {\bibinfo {volume} {87}},\ \bibinfo
  {pages} {023201} (\bibinfo {year} {2013})}\BibitemShut {NoStop}%
\bibitem [{\citenamefont {Tomita}\ \emph {et~al.}(2014)\citenamefont {Tomita},
  \citenamefont {Sawada}, \citenamefont {Porokhnyuk},\ and\ \citenamefont
  {Ueda}}]{Tomita2014}%
  \BibitemOpen
  \bibfield  {author} {\bibinfo {author} {\bibfnamefont {S.}~\bibnamefont
  {Tomita}}, \bibinfo {author} {\bibfnamefont {K.}~\bibnamefont {Sawada}},
  \bibinfo {author} {\bibfnamefont {A.}~\bibnamefont {Porokhnyuk}}, \ and\
  \bibinfo {author} {\bibfnamefont {T.}~\bibnamefont {Ueda}},\ }\href {\doibase
  10.1103/PhysRevLett.113.235501} {\bibfield  {journal} {\bibinfo  {journal}
  {Phys. Rev. Lett.}\ }\textbf {\bibinfo {volume} {113}},\ \bibinfo {pages}
  {235501} (\bibinfo {year} {2014})}\BibitemShut {NoStop}%
\bibitem [{\citenamefont {Yao}\ and\ \citenamefont {Padgett}(2011)}]{Yao2011}%
  \BibitemOpen
  \bibfield  {author} {\bibinfo {author} {\bibfnamefont {A.}~\bibnamefont
  {Yao}}\ and\ \bibinfo {author} {\bibfnamefont {M.}~\bibnamefont {Padgett}},\
  }\href {\doibase 10.1364/AOP.3.000161} {\bibfield  {journal} {\bibinfo
  {journal} {Advances in Optics and Photonics}\ }\textbf {\bibinfo {volume}
  {3}},\ \bibinfo {pages} {161} (\bibinfo {year} {2011})}\BibitemShut {NoStop}%
\bibitem [{\citenamefont {Bliokh}\ and\ \citenamefont
  {Nori}(2015)}]{Bliokh2015}%
  \BibitemOpen
  \bibfield  {author} {\bibinfo {author} {\bibfnamefont {K.~Y.}\ \bibnamefont
  {Bliokh}}\ and\ \bibinfo {author} {\bibfnamefont {F.}~\bibnamefont {Nori}},\
  }\href {\doibase http://dx.doi.org/10.1016/j.physrep.2015.06.003} {\bibfield
  {journal} {\bibinfo  {journal} {Phys. Rep.}\ }\textbf {\bibinfo {volume}
  {592}},\ \bibinfo {pages} {1 } (\bibinfo {year} {2015})}\BibitemShut
  {NoStop}%
\bibitem [{\citenamefont {Afanasiev}\ and\ \citenamefont
  {Stepanovsky}(1996)}]{Afanasiev2007}%
  \BibitemOpen
  \bibfield  {author} {\bibinfo {author} {\bibfnamefont {G.~N.}\ \bibnamefont
  {Afanasiev}}\ and\ \bibinfo {author} {\bibfnamefont {Y.~P.}\ \bibnamefont
  {Stepanovsky}},\ }\href {\doibase 10.1007/BF02731014} {\bibfield  {journal}
  {\bibinfo  {journal} {Nuovo Cimento A}\ }\textbf {\bibinfo {volume} {109}},\
  \bibinfo {pages} {271} (\bibinfo {year} {1996})}\BibitemShut {NoStop}%
\bibitem [{\citenamefont {Bliokh}\ and\ \citenamefont
  {Nori}(2011)}]{Bliokh2011}%
  \BibitemOpen
  \bibfield  {author} {\bibinfo {author} {\bibfnamefont {K.~Y.}\ \bibnamefont
  {Bliokh}}\ and\ \bibinfo {author} {\bibfnamefont {F.}~\bibnamefont {Nori}},\
  }\href {\doibase 10.1103/PhysRevA.83.021803} {\bibfield  {journal} {\bibinfo
  {journal} {Phys. Rev. A}\ }\textbf {\bibinfo {volume} {83}},\ \bibinfo
  {pages} {021803} (\bibinfo {year} {2011})}\BibitemShut {NoStop}%
\bibitem [{\citenamefont {Coles}\ and\ \citenamefont
  {Andrews}(2012)}]{Coles2012}%
  \BibitemOpen
  \bibfield  {author} {\bibinfo {author} {\bibfnamefont {M.~M.}\ \bibnamefont
  {Coles}}\ and\ \bibinfo {author} {\bibfnamefont {D.~L.}\ \bibnamefont
  {Andrews}},\ }\href {\doibase 10.1103/PhysRevA.85.063810} {\bibfield
  {journal} {\bibinfo  {journal} {Phys. Rev. A}\ }\textbf {\bibinfo {volume}
  {85}},\ \bibinfo {pages} {063810} (\bibinfo {year} {2012})}\BibitemShut
  {NoStop}%
\bibitem [{\citenamefont {Cameron}\ \emph {et~al.}(2012)\citenamefont
  {Cameron}, \citenamefont {Barnett},\ and\ \citenamefont
  {Yao}}]{Cameron2012a}%
  \BibitemOpen
  \bibfield  {author} {\bibinfo {author} {\bibfnamefont {R.~P.}\ \bibnamefont
  {Cameron}}, \bibinfo {author} {\bibfnamefont {S.~M.}\ \bibnamefont
  {Barnett}}, \ and\ \bibinfo {author} {\bibfnamefont {A.~M.}\ \bibnamefont
  {Yao}},\ }\href {http://stacks.iop.org/1367-2630/14/i=5/a=053050} {\bibfield
  {journal} {\bibinfo  {journal} {New J. of Phys.}\ }\textbf {\bibinfo {volume}
  {14}},\ \bibinfo {pages} {053050} (\bibinfo {year} {2012})}\BibitemShut
  {NoStop}%
\bibitem [{\citenamefont {Barnett}\ \emph {et~al.}(2012)\citenamefont
  {Barnett}, \citenamefont {Cameron},\ and\ \citenamefont {Yao}}]{Barnett2012}%
  \BibitemOpen
  \bibfield  {author} {\bibinfo {author} {\bibfnamefont {S.~M.}\ \bibnamefont
  {Barnett}}, \bibinfo {author} {\bibfnamefont {R.~P.}\ \bibnamefont
  {Cameron}}, \ and\ \bibinfo {author} {\bibfnamefont {A.~M.}\ \bibnamefont
  {Yao}},\ }\href {\doibase 10.1103/PhysRevA.86.013845} {\bibfield  {journal}
  {\bibinfo  {journal} {Phys. Rev. A}\ }\textbf {\bibinfo {volume} {86}},\
  \bibinfo {pages} {013845} (\bibinfo {year} {2012})}\BibitemShut {NoStop}%
\bibitem [{\citenamefont {Bliokh}\ \emph
  {et~al.}(2014{\natexlab{a}})\citenamefont {Bliokh}, \citenamefont {Kivshar},\
  and\ \citenamefont {Nori}}]{Bliokh2014}%
  \BibitemOpen
  \bibfield  {author} {\bibinfo {author} {\bibfnamefont {K.~Y.}\ \bibnamefont
  {Bliokh}}, \bibinfo {author} {\bibfnamefont {Y.~S.}\ \bibnamefont {Kivshar}},
  \ and\ \bibinfo {author} {\bibfnamefont {F.}~\bibnamefont {Nori}},\ }\href
  {\doibase 10.1103/PhysRevLett.113.033601} {\bibfield  {journal} {\bibinfo
  {journal} {Phys. Rev. Lett.}\ }\textbf {\bibinfo {volume} {113}},\ \bibinfo
  {pages} {033601} (\bibinfo {year} {2014}{\natexlab{a}})}\BibitemShut
  {NoStop}%
\bibitem [{\citenamefont {Bliokh}\ \emph
  {et~al.}(2014{\natexlab{b}})\citenamefont {Bliokh}, \citenamefont {Dressel},\
  and\ \citenamefont {Nori}}]{Bliokh2014a}%
  \BibitemOpen
  \bibfield  {author} {\bibinfo {author} {\bibfnamefont {K.~Y.}\ \bibnamefont
  {Bliokh}}, \bibinfo {author} {\bibfnamefont {J.}~\bibnamefont {Dressel}}, \
  and\ \bibinfo {author} {\bibfnamefont {F.}~\bibnamefont {Nori}},\ }\href
  {http://stacks.iop.org/1367-2630/16/i=9/a=093037} {\bibfield  {journal}
  {\bibinfo  {journal} {New Journal of Physics}\ }\textbf {\bibinfo {volume}
  {16}},\ \bibinfo {pages} {093037} (\bibinfo {year}
  {2014}{\natexlab{b}})}\BibitemShut {NoStop}%
\bibitem [{\citenamefont {Bliokh}\ \emph {et~al.}(2015)\citenamefont {Bliokh},
  \citenamefont {Rodr{\'\i}guez-Fortu{\~n}o}, \citenamefont {Nori},\ and\
  \citenamefont {Zayats}}]{Bliokh2015a}%
  \BibitemOpen
  \bibfield  {author} {\bibinfo {author} {\bibfnamefont {K.~Y.}\ \bibnamefont
  {Bliokh}}, \bibinfo {author} {\bibfnamefont {F.}~\bibnamefont
  {Rodr{\'\i}guez-Fortu{\~n}o}}, \bibinfo {author} {\bibfnamefont
  {F.}~\bibnamefont {Nori}}, \ and\ \bibinfo {author} {\bibfnamefont {A.~V.}\
  \bibnamefont {Zayats}},\ }\href@noop {} {\bibfield  {journal} {\bibinfo
  {journal} {Nature Photonics}\ }\textbf {\bibinfo {volume} {9}},\ \bibinfo
  {pages} {796} (\bibinfo {year} {2015})}\BibitemShut {NoStop}%
\bibitem [{\citenamefont {Bliokh}\ \emph {et~al.}(2016)\citenamefont {Bliokh},
  \citenamefont {Bekshaev},\ and\ \citenamefont {Nori}}]{Bliokh2016}%
  \BibitemOpen
  \bibfield  {author} {\bibinfo {author} {\bibfnamefont {K.~Y.}\ \bibnamefont
  {Bliokh}}, \bibinfo {author} {\bibfnamefont {A.~Y.}\ \bibnamefont
  {Bekshaev}}, \ and\ \bibinfo {author} {\bibfnamefont {F.}~\bibnamefont
  {Nori}},\ }\href@noop {} {\bibfield  {journal} {\bibinfo  {journal} {New
  Journal of Physics}\ }\textbf {\bibinfo {volume} {18}},\ \bibinfo {pages}
  {089503} (\bibinfo {year} {2016})}\BibitemShut {NoStop}%
\bibitem [{\citenamefont {Dressel}\ \emph {et~al.}(2015)\citenamefont
  {Dressel}, \citenamefont {Bliokh},\ and\ \citenamefont {Nori}}]{Dressel2015}%
  \BibitemOpen
  \bibfield  {author} {\bibinfo {author} {\bibfnamefont {J.}~\bibnamefont
  {Dressel}}, \bibinfo {author} {\bibfnamefont {K.~Y.}\ \bibnamefont {Bliokh}},
  \ and\ \bibinfo {author} {\bibfnamefont {F.}~\bibnamefont {Nori}},\ }\href
  {\doibase http://dx.doi.org/10.1016/j.physrep.2015.06.001} {\bibfield
  {journal} {\bibinfo  {journal} {Physics Reports}\ }\textbf {\bibinfo {volume}
  {589}},\ \bibinfo {pages} {1 } (\bibinfo {year} {2015})}\BibitemShut
  {NoStop}%
\bibitem [{\citenamefont {Ragusa}(1994)}]{Ragusa1994}%
  \BibitemOpen
  \bibfield  {author} {\bibinfo {author} {\bibfnamefont {S.}~\bibnamefont
  {Ragusa}},\ }\href {http://stacks.iop.org/0305-4470/27/i=8/a=024} {\bibfield
  {journal} {\bibinfo  {journal} {J. Phys. A: Math. Gen.}\ }\textbf {\bibinfo
  {volume} {27}},\ \bibinfo {pages} {2887} (\bibinfo {year}
  {1994})}\BibitemShut {NoStop}%
\bibitem [{\citenamefont {Ragusa}(1996)}]{Ragusa1996}%
  \BibitemOpen
  \bibfield  {author} {\bibinfo {author} {\bibfnamefont {S.}~\bibnamefont
  {Ragusa}},\ }\href@noop {} {\bibfield  {journal} {\bibinfo  {journal}
  {Brazilian Journal of Physics}\ }\textbf {\bibinfo {volume} {26}},\ \bibinfo
  {pages} {411} (\bibinfo {year} {1996})}\BibitemShut {NoStop}%
\bibitem [{\citenamefont {Fernandez-Corbaton}\ and\ \citenamefont
  {Molina-Terriza}(2013)}]{Fernandez2013}%
  \BibitemOpen
  \bibfield  {author} {\bibinfo {author} {\bibfnamefont {I.}~\bibnamefont
  {Fernandez-Corbaton}}\ and\ \bibinfo {author} {\bibfnamefont
  {G.}~\bibnamefont {Molina-Terriza}},\ }\href {\doibase
  10.1103/PhysRevB.88.085111} {\bibfield  {journal} {\bibinfo  {journal} {Phys.
  Rev. B}\ }\textbf {\bibinfo {volume} {88}},\ \bibinfo {pages} {085111}
  (\bibinfo {year} {2013})}\BibitemShut {NoStop}%
\bibitem [{\citenamefont {Fernandez-Corbaton}(2013)}]{Fernandez2013a}%
  \BibitemOpen
  \bibfield  {author} {\bibinfo {author} {\bibfnamefont {I.}~\bibnamefont
  {Fernandez-Corbaton}},\ }\href@noop {} {\bibfield  {journal} {\bibinfo
  {journal} {Optics express}\ }\textbf {\bibinfo {volume} {21}},\ \bibinfo
  {pages} {29885} (\bibinfo {year} {2013})}\BibitemShut {NoStop}%
\bibitem [{\citenamefont {Robinson}(2012)}]{Robinson2012}%
  \BibitemOpen
  \bibfield  {author} {\bibinfo {author} {\bibfnamefont {D.}~\bibnamefont
  {Robinson}},\ }\href@noop {} {\emph {\bibinfo {title} {A Course in the Theory
  of Groups}}},\ Vol.~\bibinfo {volume} {80}\ (\bibinfo  {publisher} {Springer
  Science \& Business Media},\ \bibinfo {year} {2012})\BibitemShut {NoStop}%
\bibitem [{\citenamefont {Berry}(2009)}]{Berry2009}%
  \BibitemOpen
  \bibfield  {author} {\bibinfo {author} {\bibfnamefont {M.~V.}\ \bibnamefont
  {Berry}},\ }\href {http://stacks.iop.org/1464-4258/11/i=9/a=094001}
  {\bibfield  {journal} {\bibinfo  {journal} {J. Opt A: Pure Appl. Opt.}\
  }\textbf {\bibinfo {volume} {11}},\ \bibinfo {pages} {094001} (\bibinfo
  {year} {2009})}\BibitemShut {NoStop}%
\bibitem [{\citenamefont {Claborn}\ \emph {et~al.}(2008)\citenamefont
  {Claborn}, \citenamefont {Isborn}, \citenamefont {Kaminsky},\ and\
  \citenamefont {Kahr}}]{Claborn2008}%
  \BibitemOpen
  \bibfield  {author} {\bibinfo {author} {\bibfnamefont {K.}~\bibnamefont
  {Claborn}}, \bibinfo {author} {\bibfnamefont {C.}~\bibnamefont {Isborn}},
  \bibinfo {author} {\bibfnamefont {W.}~\bibnamefont {Kaminsky}}, \ and\
  \bibinfo {author} {\bibfnamefont {B.}~\bibnamefont {Kahr}},\ }\href {\doibase
  10.1002/anie.200704559} {\bibfield  {journal} {\bibinfo  {journal}
  {Angewandte Chemie International Edition}\ }\textbf {\bibinfo {volume}
  {47}},\ \bibinfo {pages} {5706} (\bibinfo {year} {2008})}\BibitemShut
  {NoStop}%
\bibitem [{\citenamefont {Condon}(1937)}]{Condon1937}%
  \BibitemOpen
  \bibfield  {author} {\bibinfo {author} {\bibfnamefont {E.~U.}\ \bibnamefont
  {Condon}},\ }\href {\doibase 10.1103/RevModPhys.9.432} {\bibfield  {journal}
  {\bibinfo  {journal} {Rev. Mod. Phys.}\ }\textbf {\bibinfo {volume} {9}},\
  \bibinfo {pages} {432} (\bibinfo {year} {1937})}\BibitemShut {NoStop}%
\bibitem [{\citenamefont {Born}(1972)}]{Born1972}%
  \BibitemOpen
  \bibfield  {author} {\bibinfo {author} {\bibfnamefont {M.}~\bibnamefont
  {Born}},\ }\href@noop {} {\emph {\bibinfo {title} {Optik Springer-Verlag}}}\
  (\bibinfo  {publisher} {Berlin},\ \bibinfo {year} {1972})\BibitemShut
  {NoStop}%
\bibitem [{\citenamefont {Fedorov}(1959)}]{Fedorov1959}%
  \BibitemOpen
  \bibfield  {author} {\bibinfo {author} {\bibfnamefont {F.}~\bibnamefont
  {Fedorov}},\ }\href@noop {} {\bibfield  {journal} {\bibinfo  {journal} {Opt.
  Spectrosc.}\ }\textbf {\bibinfo {volume} {6}},\ \bibinfo {pages} {49}
  (\bibinfo {year} {1959})}\BibitemShut {NoStop}%
\bibitem [{\citenamefont {Fedorov}(1976)}]{Fedorov1976}%
  \BibitemOpen
  \bibfield  {author} {\bibinfo {author} {\bibfnamefont {F.~I.}\ \bibnamefont
  {Fedorov}},\ }\href@noop {} {\emph {\bibinfo {title} {Teorija girotropii}}}\
  (\bibinfo  {publisher} {Izd. Nauka i Technika},\ \bibinfo {year}
  {1976})\BibitemShut {NoStop}%
\bibitem [{\citenamefont {Post}(1962)}]{Post1962}%
  \BibitemOpen
  \bibfield  {author} {\bibinfo {author} {\bibfnamefont {E.~J.}\ \bibnamefont
  {Post}},\ }\href@noop {} {\emph {\bibinfo {title} {Formal structure of
  electromagnetics: general covariance and electromagnetics}}}\ (\bibinfo
  {publisher} {North-Holland Publishing Company},\ \bibinfo {address}
  {Amsterdam},\ \bibinfo {year} {1962})\BibitemShut {NoStop}%
\bibitem [{\citenamefont {Rado}\ and\ \citenamefont {Folen}(1962)}]{Rado1962}%
  \BibitemOpen
  \bibfield  {author} {\bibinfo {author} {\bibfnamefont {G.~T.}\ \bibnamefont
  {Rado}}\ and\ \bibinfo {author} {\bibfnamefont {V.~J.}\ \bibnamefont
  {Folen}},\ }\href {\doibase http://dx.doi.org/10.1063/1.1728630} {\bibfield
  {journal} {\bibinfo  {journal} {J. Appl. Phys.}\ }\textbf {\bibinfo {volume}
  {33}},\ \bibinfo {pages} {1126} (\bibinfo {year} {1962})}\BibitemShut
  {NoStop}%
\bibitem [{\citenamefont {Jaggard}\ \emph {et~al.}(1979)\citenamefont
  {Jaggard}, \citenamefont {Mickelson},\ and\ \citenamefont
  {Papas}}]{Jaggard1979}%
  \BibitemOpen
  \bibfield  {author} {\bibinfo {author} {\bibfnamefont {D.}~\bibnamefont
  {Jaggard}}, \bibinfo {author} {\bibfnamefont {A.}~\bibnamefont {Mickelson}},
  \ and\ \bibinfo {author} {\bibfnamefont {C.}~\bibnamefont {Papas}},\
  }\href@noop {} {\bibfield  {journal} {\bibinfo  {journal} {Appl. Phys.}\
  }\textbf {\bibinfo {volume} {18}},\ \bibinfo {pages} {211} (\bibinfo {year}
  {1979})}\BibitemShut {NoStop}%
\bibitem [{\citenamefont {Hornreich}\ and\ \citenamefont
  {Shtrikman}(1968)}]{Hornreich1968}%
  \BibitemOpen
  \bibfield  {author} {\bibinfo {author} {\bibfnamefont {R.~M.}\ \bibnamefont
  {Hornreich}}\ and\ \bibinfo {author} {\bibfnamefont {S.}~\bibnamefont
  {Shtrikman}},\ }\href {\doibase 10.1103/PhysRev.171.1065} {\bibfield
  {journal} {\bibinfo  {journal} {Phys. Rev.}\ }\textbf {\bibinfo {volume}
  {171}},\ \bibinfo {pages} {1065} (\bibinfo {year} {1968})}\BibitemShut
  {NoStop}%
\bibitem [{\citenamefont {Hornreich}(1968)}]{Hornreich1968a}%
  \BibitemOpen
  \bibfield  {author} {\bibinfo {author} {\bibfnamefont {R.~M.}\ \bibnamefont
  {Hornreich}},\ }\href {\doibase http://dx.doi.org/10.1063/1.2163466}
  {\bibfield  {journal} {\bibinfo  {journal} {J. Appl. Phys.}\ }\textbf
  {\bibinfo {volume} {39}},\ \bibinfo {pages} {432} (\bibinfo {year}
  {1968})}\BibitemShut {NoStop}%
\bibitem [{\citenamefont {Lekner}(1996)}]{Lekner1996}%
  \BibitemOpen
  \bibfield  {author} {\bibinfo {author} {\bibfnamefont {J.}~\bibnamefont
  {Lekner}},\ }\href@noop {} {\bibfield  {journal} {\bibinfo  {journal} {Pure
  and Applied Optics: Journal of the European Optical Society Part A}\ }\textbf
  {\bibinfo {volume} {5}},\ \bibinfo {pages} {417} (\bibinfo {year}
  {1996})}\BibitemShut {NoStop}%
\bibitem [{\citenamefont {Fisanov}(2007)}]{Firsanov2007}%
  \BibitemOpen
  \bibfield  {author} {\bibinfo {author} {\bibfnamefont {V.~V.}\ \bibnamefont
  {Fisanov}},\ }\href {\doibase 10.1134/S1064226907090094} {\bibfield
  {journal} {\bibinfo  {journal} {Journal of Communications Technology and
  Electronics}\ }\textbf {\bibinfo {volume} {52}},\ \bibinfo {pages} {1006}
  (\bibinfo {year} {2007})}\BibitemShut {NoStop}%
\bibitem [{\citenamefont {{Cho}}(2015)}]{Cho2015}%
  \BibitemOpen
  \bibfield  {author} {\bibinfo {author} {\bibfnamefont {K.}~\bibnamefont
  {{Cho}}},\ }\href@noop {} {\bibfield  {journal} {\bibinfo  {journal} {ArXiv
  e-prints}\ } (\bibinfo {year} {2015})},\ \Eprint
  {http://arxiv.org/abs/1501.01078} {arXiv:1501.01078} \BibitemShut {NoStop}%
\bibitem [{\citenamefont {Bialynicki-Birula}(1996)}]{Bialynicki1996}%
  \BibitemOpen
  \bibfield  {author} {\bibinfo {author} {\bibfnamefont {I.}~\bibnamefont
  {Bialynicki-Birula}},\ }in\ \href@noop {} {\emph {\bibinfo {booktitle}
  {Coherence and Quantum Optics VII}}}\ (\bibinfo  {publisher} {Springer},\
  \bibinfo {year} {1996})\ pp.\ \bibinfo {pages} {313--322}\BibitemShut
  {NoStop}%
\bibitem [{Note1()}]{Note1}%
  \BibitemOpen
  \bibinfo {note} {If the relation $ \mu (\protect \bm {r})/\varepsilon
  (\protect \bm {r}) = \protect \mathrm {Const} $ holds in medium with
  constituent relations in Eqs.~(\ref {eq:Fedorov}--\ref {eq:Pos1}), this
  medium is self-dual according to Ref.~\cite {Fernandez2013}. In this case, we
  can also rewrite Eq.~(\ref {eq:drho}) in the form of conservation law for
  redefined $ \rho _\chi $ and $ \protect \bm {J}_\chi $}\BibitemShut {NoStop}%
\bibitem [{\citenamefont {Hu}\ \emph {et~al.}(1987)\citenamefont {Hu},
  \citenamefont {Kattawar}, \citenamefont {Parkin},\ and\ \citenamefont
  {Herb}}]{Hu1987}%
  \BibitemOpen
  \bibfield  {author} {\bibinfo {author} {\bibfnamefont {C.-R.}\ \bibnamefont
  {Hu}}, \bibinfo {author} {\bibfnamefont {G.~W.}\ \bibnamefont {Kattawar}},
  \bibinfo {author} {\bibfnamefont {M.~E.}\ \bibnamefont {Parkin}}, \ and\
  \bibinfo {author} {\bibfnamefont {P.}~\bibnamefont {Herb}},\ }\href@noop {}
  {\bibfield  {journal} {\bibinfo  {journal} {Applied optics}\ }\textbf
  {\bibinfo {volume} {26}},\ \bibinfo {pages} {4159} (\bibinfo {year}
  {1987})}\BibitemShut {NoStop}%
\bibitem [{\citenamefont {Glazer}\ and\ \citenamefont
  {Burns}(2013)}]{Glazer2013}%
  \BibitemOpen
  \bibfield  {author} {\bibinfo {author} {\bibfnamefont {M.}~\bibnamefont
  {Glazer}}\ and\ \bibinfo {author} {\bibfnamefont {G.}~\bibnamefont {Burns}},\
  }\href {https://books.google.co.jp/books?id=\_mww5BoUn\_wC} {\emph {\bibinfo
  {title} {Space Groups for Solid State Scientists}}}\ (\bibinfo  {publisher}
  {Elsevier Science},\ \bibinfo {year} {2013})\BibitemShut {NoStop}%
\bibitem [{Note2()}]{Note2}%
  \BibitemOpen
  \bibinfo {note} {Here, under the helicity we mean the projection of the spin
  onto the direction of propagation, $(\protect \bm {\protect \mathcal {S}}
  \cdot \protect \mathaccentV {tilde}07E{\protect \bm {p}})$. We note that to
  discuss the physical helicity expressed through the difference in population
  of left and right polarized photons, one has to construct a photon wave
  function in chiral crystals with $S_{4}$ and $D_{2d}$ groups, which is,
  however, beyond the scope of our symmetry analysis}\BibitemShut {NoStop}%
\end{thebibliography}%

\end{document}